\documentclass[runningheads]{llncs}

\usepackage[utf8]{inputenc}
\usepackage[T1]{fontenc}
\usepackage{graphicx}
\usepackage{xspace}
\usepackage{stmaryrd}
\usepackage{amsmath,amssymb}
\usepackage{latexsym} 
\usepackage{hyperref}

\newcommand{\languageName}[1]{\text{#1}\xspace}
\newcommand{\rest}{\languageName{REST}}
\newcommand{\saferestscriptlang}{\languageName{SRS}}
\newcommand{\typescriptlang}{\languageName{TypeScript}}
\newcommand{\boogielang}{\languageName{Boogie}}
\newcommand{\headrestlang}{\languageName{HeadREST}}
\newcommand{\whileylang}{\languageName{Whiley}}
\newcommand{\dafnylang}{\languageName{Dafny}}
\newcommand{\javascriptlang}{\languageName{JavaScript}}
\newcommand{\whiplang}{\languageName{Whip}}

\newcommand{\LangHR}{\headrestlang} 
\newcommand{\LangSRS}{\saferestscriptlang} 

\newcommand{\headrest}{\lstinline[language=headrest]}
\newcommand{\safescript}{\lstinline[language=safescript]}
\newcommand{\boogie}{\lstinline[language=boogie]}
\newcommand{\javascript}{\lstinline[language=javascript]}
\newcommand{\saferestscript}{\lstinline[language=saferestscript]}

\newcommand{\keyword}[1]{\mathsf{#1}}
\newcommand{\booleank}{\keyword{Boolean}}
\newcommand{\integerk}{\keyword{Integer}}
\newcommand{\stringk}{\keyword{String}}
\newcommand{\naturalk}{\keyword{Natural}}
\newcommand{\regexpk}{\keyword{Regexp}}
\newcommand{\uritemplatek}{\keyword{URITemplate}}
\newcommand{\Responsek}{\keyword{Response}}
\newcommand{\Requestk}{\keyword{Request}}
\newcommand{\emptyk}{\keyword{Empty}}
\newcommand{\anyk}{\keyword{Any}}

\newcommand{\vark}{\keyword{var}}
\newcommand{\typek}{\keyword{type}}

\newcommand{\truek}{\keyword{true}}
\newcommand{\falsek}{\keyword{false}}

\newcommand{\uritk}{\keyword{URITemplate}}

\newcommand{\getk}{\keyword{get}}
\newcommand{\putk}{\keyword{put}}
\newcommand{\postk}{\keyword{post}}
\newcommand{\deletek}{\keyword{delete}}

\newcommand{\ink}{\keyword{in}}
\newcommand{\isdefinedk}{\keyword{isdefined}}
\newcommand{\lengthk}{\keyword{length}}
\newcommand{\sizek}{\keyword{size}}

\newcommand{\wherek}{\keyword{where}}
\newcommand{\nullk}{\keyword{null}}

\newcommand{\ifk}{\keyword{if}}
\newcommand{\thenk}{\keyword{then}}
\newcommand{\elsek}{\keyword{else}}

\newcommand{\forallk}{\keyword{forall}}
\newcommand{\existsk}{\keyword{exists}}

\newcommand{\whilek}{\keyword{while}}
\newcommand{\returnk}{\keyword{return}}

\newcommand{\assertk}{\keyword{assert}}

\newcommand{\invk}{\keyword{inv}}

\newcommand{\undefinedk}{\keyword{undefined}}

\newcommand{\ofk}{\keyword{of}}

\newcommand{\mkarrayk}{\keyword{mkarray}}
\newcommand{\voidk}{\keyword{void}}

\newcommand{\intk}{\keyword{int}}
\newcommand{\asynck}{\keyword{async}}
\newcommand{\awaitk}{\keyword{await}}

\newcommand{\specificationk}{\keyword{specification}}

\newcommand{\responseid}{\keyword{response}}
\newcommand{\requestid}{\keyword{request}}
\newcommand{\templateid}{\keyword{template}}
\newcommand{\rootid}{\keyword{root}}

\newcommand{\arraytype}[1][T]{#1\textsf{\upshape[]}}
\newcommand{\objecttype}[1][T]{\{l \colon {#1}\}}
\newcommand{\wheret}[3]{\exptype{{#1}\colon{#2}}{#3}}
\newcommand{\exptype}[2]{{#1}\:\wherek\:{#2}}
\newcommand{\reftype}{(\wheret xTe)}
\newcommand{\isin}[2]{{#1}\;\ink\;{#2}}
\newcommand{\inexp}{\isin eT}

\newcommand{\AndT}{\;\&\;} 
\newcommand{\OrE}{\;|\;} 
\newcommand{\AndE}{\;\&\;} 
\newcommand{\AndC}{\;\&\&\;} 

\newcommand{\cond}\ite

\newcommand{\Forall}[3]{\forallk\:{#1}\colon{#2}.{#3}}
\newcommand{\Exists}[3]{\existsk\:{#1}\colon{#2}.{#3}}

\newcommand{\ite}[3]{{#1}\,?\,{#2}:{#3}} 

\newcommand{\fv}{\operatorname{fv}}

\newcommand{\triple}[4]{\{{#1}\}{#2}\,{#3}\{{#4}\}}

\newcommand{\importdecl}[2]{\specificationk\,{#1}\,\ofk\,{#2}}
\newcommand{\vardecl}[3]{\vark\,{#1}\,{#2}={#3}}
\newcommand{\typedecl}[2]{\typek\,{#1}={#2}}
\newcommand{\fundecl}[4]{\asynck^? {#1}\,{#2}\,({#3})\,\{{#4}\}}

\newcommand{\ifs}[3]{\ifk\,({#1})\,\thenk\,{#2}\,\elsek\,{#3}}
\newcommand{\whiles}[3]{\whilek\,({#1})\,\invk\,{#2}\,{#3}}
\newcommand{\returns}[1]{\returnk\,{#1}}

\newcommand{\isUri}[2]{\vdash{#1}\rightarrowtriangle{#2}}

\newcommand{\opname}[1]{\operatorname{#1}}

\newcommand{\conjDD}[1]{\opname{conj_{DD}}({#1})}
\newcommand{\conjRD}[1]{\opname{conj_{RD}}({#1})}
\newcommand{\conjRR}[1]{\opname{conj_{RR}}({#1})}
\newcommand{\normr}[1]{\opname{norm}_r({#1})}
\newcommand{\norm}[1]{\opname{norm}({#1})}

\newcommand{\F}[2]{\mathbf{F} \llbracket{#1}\rrbracket ({#2})}
\newcommand{\Fnonopt}[1]{\mathbf{F} \llbracket{#1}\rrbracket}
\newcommand{\V}[1]{\mathbf{V} \llbracket{#1}\rrbracket}

\newcommand{\valuek}{\keyword{Value}}

\newcommand{\itemsk}{\keyword{Items}}

\newcommand{\VV}[2]{\mathbf{V^*} \llbracket{#1}\rrbracket_{#2}}
\newcommand{\W}[1]{\mathbf{W} \llbracket{#1}\rrbracket}
\newcommand{\Us}[2]{\mathbf{U} \llbracket{#1=#2}\rrbracket}

\newcommand{\B}[1]{\mathbf{B} \llbracket{#1}\rrbracket}

\newcommand{\If}{\textbf{if}}
\newcommand{\Then}{\textbf{then}}
\newcommand{\Else}{\textbf{else}}

\newcommand{\Value}{\keyword{Value}}

\newcommand{\maybeOf}[1]{\keyword{maybeOf}({#1})}
\newcommand{\isInt}[1]{\keyword{isInt}({#1})}
\newcommand{\toInt}[1]{\keyword{toInt}({#1})}
\newcommand{\fromInt}[1]{\keyword{fromInt}({#1})}
\newcommand{\isBool}[1]{\keyword{isBool}({#1})}

\newcommand{\fromBool}[1]{\keyword{fromBool}({#1})}
\newcommand{\true}{\keyword{True}}
\newcommand{\false}{\keyword{False}}
\newcommand{\isArray}[1]{\keyword{isArray}({#1})}

\newcommand{\isValidIndex}[2]{\keyword{isValidIndex}({#1},{#2})}
\newcommand{\getIndexValue}[2]{\keyword{getIndexValue}({#1},{#2})}
\newcommand{\arrayConst}{\keyword{arrayConst}()}
\newcommand{\arrayUpdate}[3]{\keyword{arrayUpdate}({#1},{#2},{#3})}
\newcommand{\isObject}[1]{\keyword{isObject}({#1})}

\newcommand{\hasField}[2]{\keyword{hasField}({#1},{#2})}
\newcommand{\getFieldValue}[2]{\keyword{getFieldValue}({#1},{#2})}
\newcommand{\objectConst}{\keyword{objectConst}()}
\newcommand{\objectUpdate}[3]{\keyword{objectUpdate}({#1},{#2},{#3})}
\newcommand{\isString}[1]{\keyword{isString}({#1})}

\newcommand{\fromString}[2]{\keyword{fromString}({#1},{#2})}
\newcommand{\emptyString}{\keyword{emptyString}()}

\newcommand{\nullB}{\keyword{Null}}

\newcommand{\undefinedB}{\keyword{Undefined}}

\newcommand{\restCall}[3]{\keyword{restCall}({#1},{#2},{#3})}

\newcommand{\Empty}{\epsilon}
\newcommand{\grmeq}{\; ::= \;\;}
\newcommand{\grmor}{\,\mid\,}
\newcommand{\eqdef}{\;\triangleq\;}
\newcommand{\subs}[2]{[{#1}/{#2}]}

\usepackage{xcolor}
\usepackage{etoolbox}
\usepackage{listings}
\usepackage[T1]{fontenc}
\usepackage{multicol}

\usepackage[scaled=0.8]{beramono}

\newcommand{\lstfontfamily}{\ttfamily}

\definecolor{darkviolet}{rgb}{0.5,0,0.4}
\definecolor{darkgreen}{rgb}{0,0.4,0.2} 
\definecolor{darkblue}{rgb}{0.1,0.1,0.9}
\definecolor{darkgrey}{rgb}{0.5,0.5,0.5}
\definecolor{lightblue}{rgb}{0.4,0.4,1}

\definecolor{stringColor}{rgb}{0.16,0.00,1.00}
\definecolor{annotationColor}{rgb}{0.39,0.39,0.39}
\definecolor{keywordColor}{rgb}{0.50,0.00,0.33}
\definecolor{commentColor}{rgb}{0.25,0.50,0.37}
\definecolor{lineNumberColor}{rgb}{0.47,0.47,0.47}

\lstdefinestyle{eclipse}{
  basicstyle={\lstfontfamily},
  emphstyle=\bfseries,
  keywordstyle=\color{keywordColor}\bfseries,
  commentstyle=\color{commentColor},
  stringstyle=\color{stringColor},
  numberstyle=\color{lineNumberColor}\lstfontfamily,
  showstringspaces=false,
  keepspaces = true
}

\lstdefinelanguage{typescript} {
  style=eclipse,
  keywords={break, case, catch, class, const, continue, debugger,default, delete, do, else, enum, export, extends, false, finally, for, function, if, import, in, instanceof, new, null, return, super, switch, this, throw, true, try, typeof, var, void, while, with, as, implements, interface, let, package, private, protected, public, static, yield, any, boolean, constructor, declare, get, module, require, number, set, string, symbol, type, from, of},
  morecomment=[l]{//},
  morecomment=[s]{/*}{*/},
  morestring=[b]",
  morestring=[b]'
}

\lstdefinelanguage{headrest} {
  style=eclipse,
  morekeywords={Integer, String, Boolean, Regexp, URITemplate, Any, where, Empty, Natural, Request, Response, get, put, post, delete, resource, type, var, specification, if, then, else, forall, exists, in, repof, uriof, true, false, null, isdefined, request, response, root},
  morecomment=[l]{//},
  morecomment=[s]{/*}{*/},
  morestring=[b]"
}

\lstdefinelanguage{safescript} {
  style=eclipse,
  morekeywords={Integer, Natural, Positive, String, Boolean, Any, Void, Empty, where, in, assert, type, var, if, then, else, while, inv, return, forall, exists, true, false, null, undefined, function, nat},
  morecomment=[l]{//},
  morecomment=[s]{/*}{*/},
  morestring=[b]"
}

\lstdefinelanguage{SafeRestScript} {
  language=safescript,
  morekeywords={specification, of, get, post, put, delete, async, await, synch, Request, Response},
}

\lstdefinelanguage{smtlib} {
  style=eclipse,
  alsoletter=-,
  keywords={
    as, BINARY, DECIMAL, exists, HEXADECIMAL, forall, let, match,NUMERAL, par, STRING, assert, check-sat, check-sat-assuming, declare-const, declare-datatype, declare-datatypes, declare-fun, declare-sort, define-fun, define-fun-rec, define-sort, echo, exit, get-assertions, get-assignment, get-info, get-model, get-option, get-proof, get-unsat-assumptions, get-unsat-core, get-value, pop, push, reset, reset-assertions, set-info, set-logic, set-option, true, false
  },
  morecomment=[l]{;},
}

\lstdefinelanguage{boogie} {
  style=eclipse,
  alsoletter=\#,
  keywords={assert, assume, axiom, bool, break, call, complete, const, else, ensures, exists, false, finite, forall, free, function, goto, havoc, if , implementation, int, invariant, modifies, old, procedure, requires, return, returns, true, type, unique, var, where, while},
  morecomment=[l]{//},
}

\lstdefinelanguage{JavaScript}{
  style=eclipse,
  keywords={typeof, new, true, false, catch, function, return, null, catch, switch, var, if, for, in, while, do, else, case, break, async, await},
  morecomment=[l]{//},
  morecomment=[s]{/*}{*/},
  morestring=[b]",
  morestring=[b]',
}

\begin{document}
\title{SafeRESTScript: Statically Checking REST API Consumers}
%
%
\author{
  Nuno Burnay\orcidID{0000-0001-6613-5192}
  \and
  Antónia Lopes\orcidID{0000-0003-0688-3521} 
  \and
  Vasco T. Vasconcelos\orcidID{0000-0002-9539-8861}
}
\authorrunning{Nuno Burnay, Antónia Lopes, and Vasco T.\ Vasconcelos}
%
\institute{LASIGE, Faculdade de Ciências, Universidade de Lisboa, Lisbon, Portugal}

\maketitle              

\begin{abstract} 
Consumption of REST services has become a popular means of invoking code provided by third parties, particularly in web applications.  Nowadays programmers of web applications can choose \typescriptlang over \javascriptlang to benefit from static type checking that enables validating calls to local functions or to those provided by libraries.  Errors in calls to REST services, however, can only be found at run-time.  

\smallskip

In this paper, we present SafeRESTScript (\saferestscriptlang, for short) a language that extends the support of static analysis to calls to REST services, with the ability to statically find common errors such as missing or invalid data in REST calls and misuse of the results from such calls.  SafeRESTScript features a syntax similar to \javascriptlang and
  is equipped with (i)~a rich collection of types---including objects,
  arrays and refinement types---and (ii)~primitives to natively support
  REST calls that are statically validated against specifications of
  the corresponding APIs.  Specifications are written in
  \headrestlang, a language that also features refinement types and
  supports the description of semantic aspects of REST APIs in a style
  reminiscent of Hoare triples.

\smallskip

  We present SafeRESTScript and its validation system, based on a general-purpose
  verification tool (\boogielang). The evaluation of SafeRESTScript
  and of the prototype implementations for its validator,
  available in the form of an Eclipse plugin, is also discussed.

\end{abstract}


\section{Introduction}

During the last decades web services have become an important
building block in the construction of distributed applications. \rest
web services in particular have become quite
popular~\cite{Fielding:2002,Richardson:2007}. These services, through
specific application programming interfaces, allow
consumers to access and manipulate representations of web resources,
identified by Unique Resource Identifiers, by using the operations
offered by HTTP. Nowadays a very large number of APIs offer interfaces
of \rest services~\cite{Levin:2015} and many software companies expose \rest APIs for their services (e.g., Google Gmail\footnote{\url{https://developers.google.com/gmail/api/}} and 
Microsoft Azure\footnote{\url{https://docs.microsoft.com/en-us/rest/api/azure/}}).

Since so many applications are designed to offer \rest APIs, the
consumption of \rest services has become a popular means of
invoking code provided by third parties. However, the support
available to programmers for writing code that consumes these services
is extremely limited when compared to the sort of support offered when
invoking external libraries provided by third parties.  The practical
impact of this problem is attested by a study on a large-scale payment
company which concluded that errors in invocations of \rest services,
related to invalid or missing data, cause most of the failures in
API consumer code~\cite{Aue:2018}.

The fact that programmers have no way of knowing whether their calls
to \rest APIs are correct until run-time was identified as one of the
four major research challenges for the consumption of web
APIs~\cite{Wittern:2017:2}. This state of affairs led to an
inter-procedural string analysis proposal to statically check \rest
calls in JavaScript~\cite{Wittern:2017}.
The solution checks whether a request to a service conforms to a given
API specification written in OpenAPI\footnote{\url{https://swagger.io/specification/}}, and involves
checking whether the endpoint targeted by the request is valid and the
request has the expected data, all this according to a given specification.
Since OpenAPI has severe limitations on what can be expressed about
the exchanged data, there are many errors related to invalid or
missing data in requests that cannot be addressed by this
approach. Moreover, since the models of response data are not taken
into consideration, misuse of the result to \rest calls cannot be
addressed although correctness of consumer code dealing with responses crucially depends on this.

This paper presents an approach to API consumer code development based
on two new languages: \headrestlang, a specification language for \rest
APIs with a rich type system that supports the specification of
semantic aspects of \rest APIs  in a style reminiscent of Hoare triples;
and \saferestscriptlang (short for \languageName{SafeRESTScript}), a
subset of JavaScript equipped with (i)~types and strong static analysis
and (ii)~primitives to natively support \rest calls that are statically
validated against \headrestlang specifications of the corresponding
APIs.  The \saferestscriptlang compiler generates JavaScript, making
it easy to use the two languages together and providing a solution for
the execution of \saferestscriptlang programs in many different
execution environments. The \saferestscriptlang compiler comes in the
form of an Eclipse plugin which is publicly
available for download\footnote{\url{http://rss.di.fc.ul.pt/tools/confident/}}.
No instalation is required to try \headrestlang as its validator can be exercised from a browser.

\paragraph{Structure of the paper.} The paper starts with a tour through our approach
(Section~\ref{sec:saferestscript-motivation}) and a brief introduction
to \headrestlang specification language (Section~\ref{sec:headrest}).
Then, we present the \saferestscriptlang programming language and its
translation to  \boogielang~\cite{Leino:2005}, which lies at the basis of the
validation system.
This validation system 
guarantees the detection
of errors in \rest calls as well as common runtime errors, like null
dereference, division by zero, or accesses outside arrays bounds.
(aspects that TypeScript cannot ensure in compilation time).  
Like the TypeScript compiler, the \saferestscriptlang compiler generates JavaScript. However, unlike TypeScript, \saferestscriptlang is not a JavaScript extension. The goal is not to have a full-fledge language but rather show that it is possible to improve support for writing code that consumes \rest services. 
Section~\ref{sec:evaluation} presents the evaluation of
\saferestscriptlang on a variety consumer code of different \rest APIs. 
Section~\ref{sec:relatedwork} discusses related work and
Section~\ref{sec:conclusion} concludes the paper by pointing towards
future work.

In the appendix the reader can find more details about the validation of \saferestscriptlang presented in Section~\ref{sec:saferestscript}, namely the complete set of rules for the 
\boogielang axiomatizations for \saferestscriptlang. 


\section{Overview of the Approach}
\label{sec:saferestscript-motivation}

This section presents the motivation for \saferestscriptlang and walks
through its approach by means of an example. First we show how
\headrestlang allows us to specify \rest APIs and capture properties
that are important and cannot be expressed in currently available Interface 
Description Languages (IDL) such as OpenAPI.  
Then we show how \LangSRS allows programming clients
of \rest services and how to rely on the compiler to check whether (i)
\rest calls conform to the specified service interface and (ii) the
response data is correctly used, thus avoiding run-time errors.

\subsection{Background} 

Applications that consume \rest APIs communicate with the service provider through calls to the API endpoints, that is to say, by sending requests for the execution of a HTTP method over an URL. The URL of the request identifies a web resource and additionally can provide values for some optional parameters; additional data can be sent in the request body. The service provider sends back a response that  carries, among other data, a response status code indicating whether the request has been successfully completed. 

\begin{figure}[h]
 \includegraphics[width=0.7\textwidth]{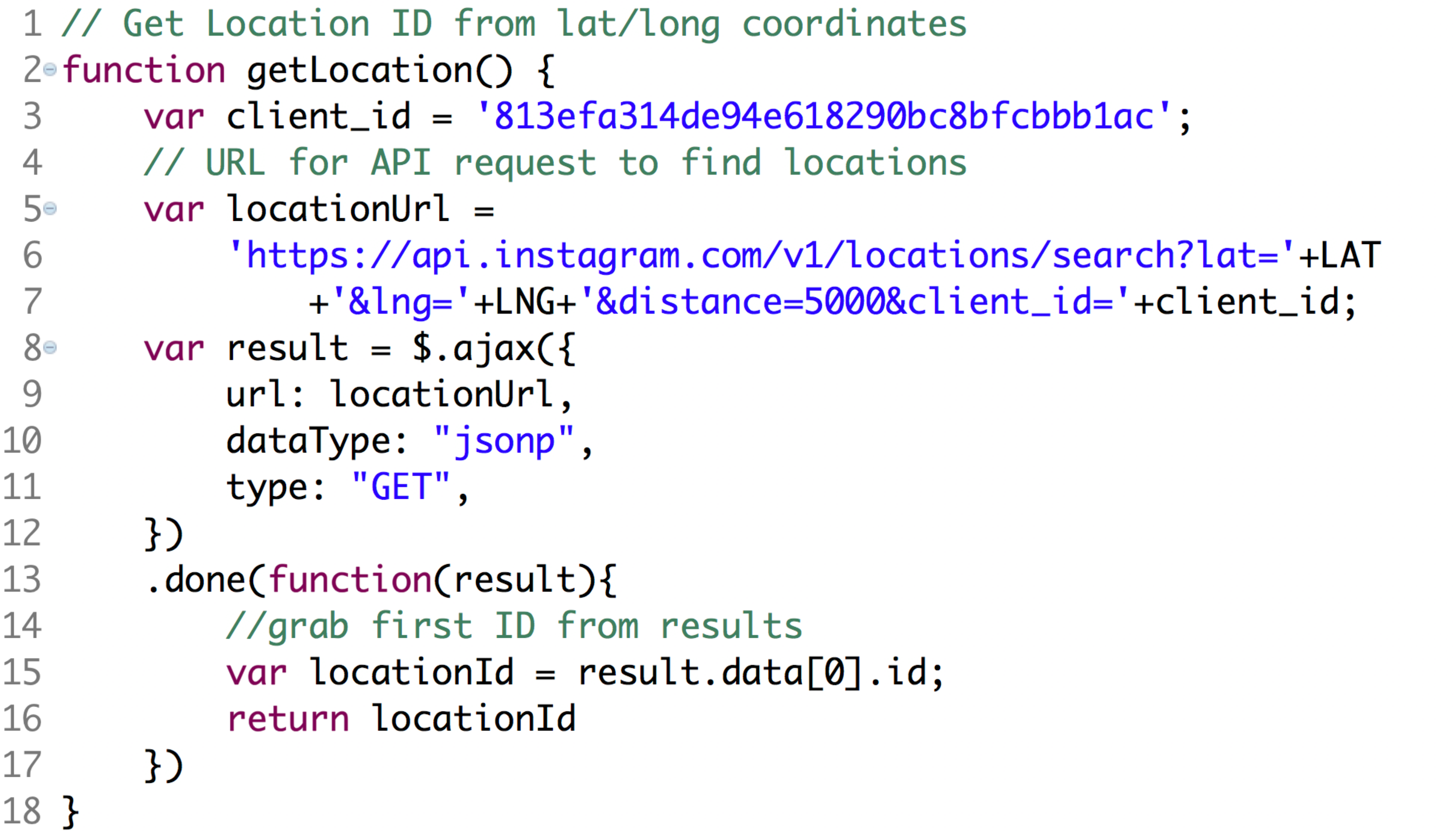}
 \caption{Excerpt of \javascriptlang code with a call to an endpoint of the Instagram API}
  \label{fig:appInstangram}
\end{figure}

Figure~\ref{fig:appInstangram} shows an excerpt of a \javascriptlang
application\footnote{\url{https://github.com/mjhea0/thinkful-mentor/blob/610695fb66852379a3549eec609962f2f351f086/frontend/instagram-search/app.js}}
that performs a call to an endpoint of the Instagram API to search for
locations by geographic coordinates. According to the API
documentation\footnote{\url{https://apiharmony-open.mybluemix.net/public/apis/instagram\#get_locations_search}},
this endpoint has several optional parameters, including
\headrest|lat|, \headrest|lng|, and \headrest|distance|, all used in the example. 
The center of the search must be defined and there are
three different ways of doing it.  Although \headrest|lat| and
\headrest|lng| are optional, if one is used, the other is also
required. The distance is optional (default value is 1000m) 
and its maximum value is 5000.  In the success case---signalled by
response code 200---the response body is an object with field
\headrest|data| containing an array of objects with field
\headrest|id|, among others.

Code that consumes this endpoint may contain different sorts of
errors. Calls may not conform to the specified interface: for instance
the request may contain a value for \headrest|lat| but not for
\headrest|lng|, or it may contain a value for \headrest|distance| that
exceeds the maximum value or simply that is not an integer. Moreover,
the response data may not be correctly used. This is the case in the
example: if the call succeeds, then line 15 accesses a possibly
non-existent element of the array in field \headrest|data|.

The fact that the model of the response data might depend 
on the provided input is an additional source of errors. For example, the
endpoint in the GitLab API to get all wiki pages for a given
project\footnote{\url{https://gitlab.com/gitlab-org/gitlab-foss/blob/swagger-api/doc/api/wikis.md}}
features an optional boolean parameter \headrest|with-content| to indicate
whether the pages’ content must be included in the response. Hence, the
response body is an array of objects that contains field
\headrest|content| only when the request has value \headrest|true| for field
\headrest|with-content|.

In order to avoid such errors, programmers need to carefully read API
documentation. The situation is worsened as this sort of
documentation tends to be vague and imprecise, even when available in
a formal document.  Limitations in the expressiveness of existing
IDLs---and in particular of OpenAPI, the de facto standard for specifying
\rest APIs---make programmers resort to natural language for conveying
extra information. In the case of the two endpoints considered here
this is in fact what happens since most of the properties under
discussion are not expressible in the IDLs used for the documentation.

Such state of affairs lead us to develop an approach that supports the
detection of common errors at compile-time by statically checking that
calls match APIs’ requirements and that data obtained in the response
is correctly used.

\subsection{\saferestscriptlang in Action}

Our approach builds on two pillars: \headrestlang, a specification
language for \rest APIs, and \LangSRS, a language with an expressive
type system for programming the code that consumes \rest
APIs. \saferestscriptlang features a syntax similar to \javascriptlang and
compiles to \javascriptlang.


\headrestlang resorts to types to express properties of states and
of data exchanged in interactions and pairs of pre and
post-conditions to express the relationship between data sent in
requests and those obtained in responses, as well as the resulting state changes.
%
Two type primitives account for its expressiveness:~refinement types, \headrest{(x:T where e)}, consisting of values of type \headrest{T} that satisfy \headrest{e};~and a predicate, \headrest{e in T}, for checking if the value of expression \headrest{e} is of type \headrest{T}.
%

The endpoints exposed by an API, and their behaviour, are specified by assertions of the form $\{\phi\} \; m \; u \; \{\psi\}$ where
$\phi$ is the pre-condition, $m$ is the HTTP method; $u$ is a URI template 
that, according to the pre-condition, expands to a valid API endpoint;
and  $\psi$ is the post-condition.

\begin{figure}[t]
 \includegraphics[width=1\textwidth]{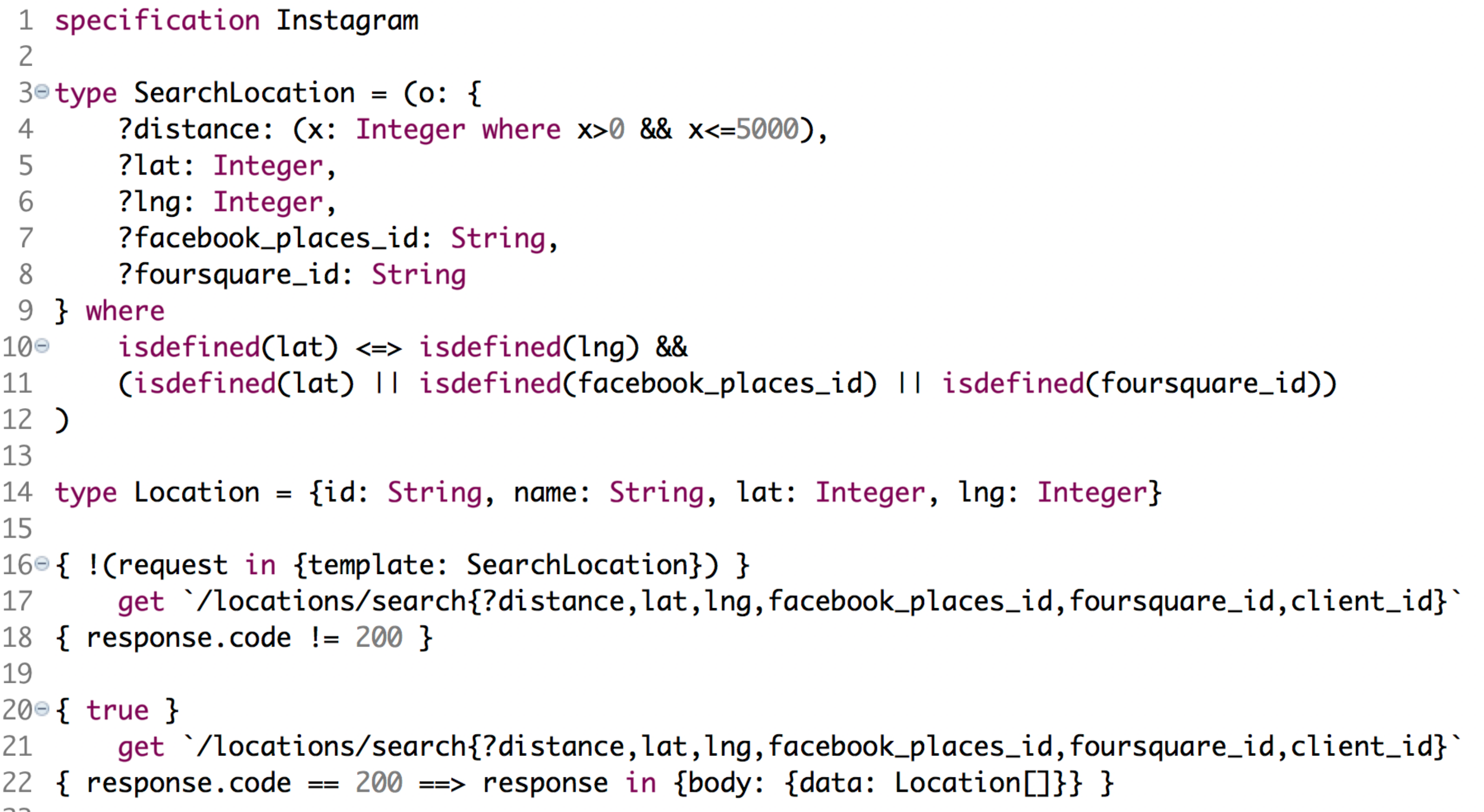}
  \caption{A \headrestlang specification for an Instagram API endpoint}
  \label{fig:InstaSpec}
\end{figure}

Figure~\ref{fig:InstaSpec} shows a specification of 
the endpoint discussed before.  It starts with the
declaration of type \headrest{SearchLocation} that represents
the search data. Note how refinement types capture the endpoint
requirements for search data discussed above; 
e.g., line 10
says that fields \headrest{lat} and \headrest{lng} must be
both present or absent; the question marks in front of these two fields
indicate that they are optional.

The behaviour of the endpoint is specified by two assertions  (lines 16--22). The
first says that, if the requirements for the search
data sent in the request are not met, then the call does not succeed
(the response code is different from \headrest{200}).
The second assertion says that, if the request is
successful (the response code is \headrest{200}), 
then the response
body consists of an array of \headrest{Location}, a type
defined in line 14.

\headrestlang also supports the specification of conditions concerning
resources, their representations and their identifiers
(see~\cite{Vasconcelos:2019} for details). For instance, in the second
triple (lines 20--22), we can specify that each \headrest{Location} in
the response is the representation of a resource that can be
individually obtained through the endpoint \headrest|GET /locations/{location-id}|. 
Since these properties do not help in avoiding errors in consumer code
(individually, clients have no control over the state of the
resources), in this paper we limit our presentation to a resource-less
version of \headrestlang.

\begin{figure}[t]
  \includegraphics[width=1\textwidth]{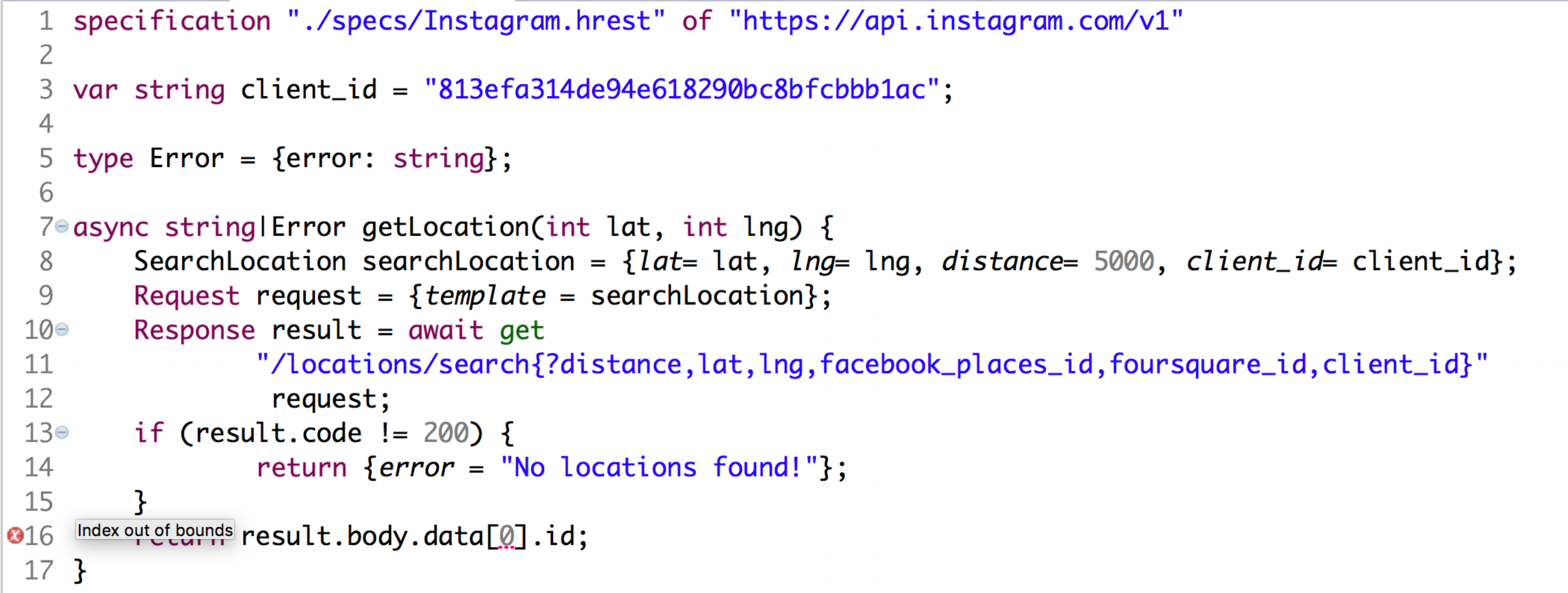}
  \\[0.5em]
  \includegraphics[width=1\textwidth]{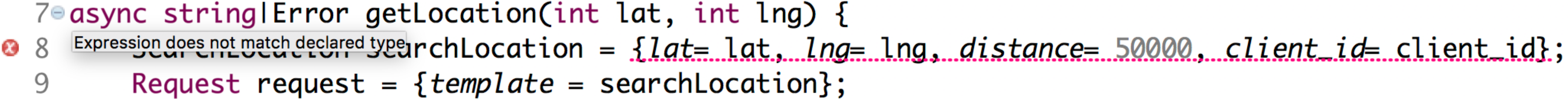}
 \caption{Example of \saferestscriptlang code consuming an Instagram API endpoint}
  \label{fig:InstaClient}
\end{figure}

Figure~\ref{fig:InstaClient} shows an \saferestscriptlang  program
similar in spirit to
the \javascriptlang code in Figure~\ref{fig:appInstangram}.
\saferestscriptlang adopts \headrestlang  types, while featuring direct
support for \rest operations.
Type checking is achieved with a translation to Boogie
and validation of the generated program with the help of Z3 SMT
solver~\cite{DeMoura:2008}.
Errors in the corresponding \boogielang code correspond to errors in
the \saferestscriptlang source.

We can see that the type checker spots an error
in the use of the response data in line 16. The specification 
ensures that the effective type of \headrest{result} in that execution
point is \headrest|{body:{data:Location[]}}| (Figure~\ref{fig:InstaSpec}, line 22). 
This means that it is safe to
access \headrest{result.body.data[i].id} only if
\headrest{i < length(result.body.data)} and, hence, line 16 
is incorrect.  Would the specification be stronger and, in line 22,
read instead 
\headrest|response.code == 200 ==>| \headrest|response in body:{data:(v: Location[] where length(v) > 0)}|, 
then the program would be valid. 
Note that the use of type
\headrest{SearchLocation} in line 8 makes sure that the data sent in
the request meets the stated necessary conditions for the request be
successful (Figure~\ref{fig:InstaSpec}, line 16). 
The figure also shows the type checker signaling an error 
if, in line 8, the value given for distance exceeds the maximum value
allowed. \saferestscriptlang further supports $\assertk$ statements that are statically validated. They are useful to check, immediately before a call to an endpoint, that a necessary
condition for the request to be successful holds. In the example, we could add
\safescript|assert(request in {template: SearchLocation})| immediately before line 10.

Valid \saferestscriptlang programs are translated to \javascriptlang,
making it easy to integrate \rest API consumer code written in
\saferestscriptlang with other \javascriptlang code, namely code of web
applications for manipulating the DOM.


\section{The \headrestlang Specification Language}
\label{sec:headrest}

\headrestlang was designed to support the specification of REST APIs
and to capture important properties that cannot be expressed in
currently available interface description languages. This section
briefly introduces the resourceless version of \headrestlang
\cite{Vasconcelos:2019}.

\begin{figure}[t]
  \centering
  \begin{align*}
  &\text{Scalar types}&
  G\grmeq & \integerk
    \grmor \stringk 
    \grmor \booleank
    \grmor \{\}
    \grmor \regexpk
    \grmor \uritk
  \\
  &\text{Types}&
  T \grmeq&
    \anyk
    \grmor G
    \grmor \objecttype
    \grmor \arraytype
    \grmor \reftype
  \\
  &\text{Constants}&
   c \grmeq &
              n 
  \grmor s
  \grmor \truek 
  \grmor \falsek
  \grmor \{\}
  \grmor u
  \grmor r
  \grmor \nullk 
  \\
    &\text{Expressions}&
    e \grmeq &
    x \grmor
    c \grmor
    f(e_1,\dots,e_n) \grmor 
    \ite eee \grmor
    \inexp \grmor
    \{l_1=e_1,\dots, l_n=e_n\}
    \\
    &&\grmor & e.l
    \grmor [e_1,\dots,e_n]
    \grmor e [e]
    \grmor \Forall xTe
    \grmor \Exists xTe
  \\
  & \text{Verbs}&
  m \grmeq & \getk \grmor \putk \grmor \postk  \grmor \deletek
  \\
  & \text{Declarations}&
  D \grmeq & 
  \triple emue ; D
  \grmor \Empty
  \end{align*}
  \caption{The syntax of \LangHR}
  \label{fig:hr_language_syntax}
\end{figure}



The syntax and validation system of \headrestlang are
influenced by the Dminor language~\cite{Bierman:2010}. Extensions and
adaptations to Dminor types, expressions and their respective
validation rules were adapted to address the specific needs of REST.
The syntax of \headrestlang is in Figure~\ref{fig:hr_language_syntax}.
It assumes a countable set of {identifiers} (denoted by $f$ or
$x,y,z$), a set of {constants} ($c$), a set of {labels}
($l,l_1,l_2,\dots$),
{integer literals} ($n$), {string literals} ($s$), a set of
{URI template literals} ($u$), and a set of {regular
  expression literals} ($r$).


\emph{Scalar types} include standard $\integerk$, $\stringk$,
$\booleank$, the REST-specific $\uritemplatek$ to represent a service
endpoint or a group of URI resources and $\regexpk$ for regular
expressions.
$\anyk$ is the top type. 
For \emph{types}, we additionally have arrays, refinement types,
and the singleton object type $\{l\colon T\}$.


\emph{Constants} include integer, string, and boolean literals, to
which $\nullk$ was added. The $\nullk$ value is of type $\anyk$ but
not of object types. The empty object type, $\{\}$, describes empty
objects and constitutes the super type of all objects. To inhabit
$\regexpk$ and $\uritemplatek$ types, two sorts of literals were added:
regular expressions and URI templates values. Regular expressions form
a subset of those in JavaScript. The syntax of URI Templates is
conform to RFC-6570~\cite{rfc6570}.

\emph{Expressions} include variables and constants, (primitive)
function calls, a conditional, arrays and object operations,
quantification, and the $\isin eT$ operator that allows checking whether a
given expression~$e$ belongs to type~$T$. Useful derived expressions include
\begin{align*}
  \isdefinedk(e.l)\eqdef  & \isin {e} {\{l : \anyk \}}
 &   
 e \AndC f \eqdef & e \; ? \; f : \falsek
\end{align*}
Universal and existencial quantifiers   were also included in the language.
They help in the declarative
description of arrays and are also useful for the expressiveness of refinement of types. 
 For example, the type that represents sorted integer arrays requires the use of quantifiers: 
 %
 \begin{equation*}
   \wheret  v {\integerk[]}
   {\Forall i \naturalk {\,i < \lengthk(v)-1 \Rightarrow v[i] \leq v[i+1] }}
 \end{equation*}


 Primitive operators include \headrest{length} and \headrest{size}
 determining, respectively, the length of an array and the size of a
 string;  \headrest{matches} checks whether a string matches a regular
 expression, and \headrest{expand} that, given a URI template and an
 object defining values for its parameters, performs the URI template
 expand operation as defined in RFC--6570~\cite{rfc6570}. 



Although \LangHR features a small core of types, the type language is
quite expressive due to the interplay between refinement types and the
$\ink$ predicate. A few examples of derived types follow, where $x$ is
a variable taken freshly.

\begin{footnotesize}
\vspace{-1.3 em}
\begin{align*}
  T \AndT U \eqdef & \wheret x \anyk {(\isin xT \AndE \isin xU)}
 &
 !T \eqdef & \wheret x \anyk {!(\isin xT)}
   \\
 \{?l\colon T\} \eqdef &  x \colon\{\}\,\wherek\ 
  {\isin{x}{\{l\colon\anyk\}} \Rightarrow \isin x {\{l\colon T\}}}
 & 
    \naturalk \eqdef & \wheret x \integerk {x\ge0}
\end{align*}
\end{footnotesize}
\vspace{-1.3 em}


\noindent The operator $\isin eT$ is essential for the expressiveness of the type system. The intersection, union and negation types are derived using this operator, and these types are the basis for many other derived types. The important multi-field object type can be derived thanks to the intersection type; e.g., 
$\{l \colon \{\} ,m \colon \stringk \}$
abbreviates  
$(\wheret {x} {\anyk\;} {\;(\isin {x} {\{ l\colon  \{\}}\} \; \& \; \isin {x} {\{ m\colon \stringk}\}))}$
which only uses core types.
An important derived type is the optional field type, $\{?l\colon T\}$,
asserting that if an object has a field $l$ then its type is $T$.
For example, if $e$ is an expression of type $\{?l\colon \booleank\}$, then expression
$\isin e {\{l\colon \anyk\} \AndC e.l}$
is valid since if $e$ has the field $l$ its type is $\booleank$ and
the good formation of $e.l$, of boolean type, only needs to be ensured in this case.


The \emph{specifications} consist of a collection of assertions
(triples), each of which describe part of the behavior of an
endpoint. Currently \LangHR supports the four main HTTP verbs:
$\getk$, $\postk$, $\putk$ and $\deletek$.

 On what concerns declarations, \LangHR supports type aliases:
 $\typek\; S = T$. Occurrences of the type identifier $S$ are directly
 replaced by $T$ in the rest of the specification, without the creation
 of a new type. 

For the specification of the pre- and post-conditions three variables are
added: $\requestid$ and $\responseid$ that correspond to the call and
the reply, and $\rootid$, the absolute URL of the entry point of the service.  The types
of the $\requestid$ and $\responseid$ variables are as follows.

\begin{footnotesize}
\begin{align*}
  \Requestk \eqdef& \{
           \text{location}\colon \stringk,
           ?\text{template}\colon\{\}, 
           \text{header}\colon\{\}, 
           ?\text{body}\colon\anyk
           \}
  \\
  \Responsek \eqdef& \{
           \text{code}\colon \integerk,
           \text{header}\colon\{\},
           ?\text{body}\colon\anyk
           \}
\end{align*} 
\end{footnotesize}

Algorithmic type checking is based on a bidirectional system of
inference rules, composed by two main relations: one that synthesizes
the type of a given expression and one that checks whether an
expression is of a given
type~\cite{DBLP:conf/icfp/DunfieldK13,DBLP:conf/ppdp/FerreiraP14,PierceTurner2000-toplas}. At
the intersection of these two relations lies \emph{semantic
  subtyping}, a relation that establishes that a type $T$ is subtype
of a type $U$ when all values that belong to $T$ also belong to $U$.
Types and contexts are translated into first-order logic (FOL)
formulae. 
The thus obtained FOL formulas are then evaluated using an SMT
solver. Our implementation uses Z3~\cite{DeMoura:2008}.  


\section{The \saferestscriptlang Programming Language}
\label{sec:saferestscript}

The \saferestscriptlang language, a shorthand for \textsc{SafeRestScript}, is 
a type-safe variant of \javascriptlang with direct support for \rest calls.
It was designed to be, at the syntactic level, as
close as possible to \javascriptlang, 
so that it may be easily used by \javascriptlang programmers. 
It transpiles to \javascriptlang, 
making it easy to integrate \rest API consumer code written in
\saferestscriptlang with  \javascriptlang code, namely code of web
applications for manipulating the DOM.

\subsection{Key Features}  

Compared with other typed extensions of \javascriptlang, such as
TypeScript~\cite{Bierman:2014}, the main novelty of
\saferestscriptlang is the incorporation of refinement types, the
in-type predicate and, most importantly, \rest endpoints as external functions.

More precisely, a \rest endpoint is seen as an impure, external function that receives a value of type  $\Requestk$, possibly changes a global resource set state, and then returns a result 
of type $\Responsek$; \rest calls are then just calls to such functions. 
Additional properties of these endpoints-as-functions, namely their specific return type, are inferred from the 
\headrestlang specification of the REST API endpoints.
Each triple in the specification specifies a relation  between the input (the $\requestid$)
and the output (the $\responseid$) of an endpoint: if the request meets the pre-condition, then the response meets the pos-condition.
In this way, a triple $\{\phi\} \; f \; \{\psi\}$ constrains the return type of the endpoint-as-function $f$ to be of type $\{\wheret{r}{\Requestk\ }{\ \phi \Rightarrow \psi} \}$. 
The conjunction of these types defines the return type of  $f$.
Note that endpoints-as-functions are, hence, total: they accept any input of type $\Requestk$, 
even those that do not meet the pre-condition of any
of their triples (in the vein of Hoare
Logic~\cite{Hoare:1969}, as opposed to that of Design by
Contract~\cite{DBLP:books/ph/Meyer97}). 

\javascriptlang is single threaded and, hence,
function calls that take time to execute
should ideally be executed asynchronously.
\rest calls fall into this category, so \LangSRS supports asynchronous
in addition to synchronous \rest calls.

\saferestscriptlang adopts the \headrestlang type system, not only for
its support for \rest operations, but also to provide precise static
type checking. In \saferestscriptlang, each variable is declared with a type 
that restricts the values that can be assigned to the variable.  
Each variable also features an
{\it effective type} that corresponds to the set of values the variable may have
at a given point in a program. The effective type changes with program
flow, but is necessarily a subtype of the declared type.


\begin{figure}[t]
  \centering
  \begin{align*}
    &\text{Constants}&
    c \grmeq &
    \dots
    \grmor \undefinedk
    \\
    &\text{Expressions}&
    e \grmeq &
    \dots
    \grmor \awaitk^? \, m \, u \, e
    \\
    &\text{Locations}&
    w \grmeq &
    x 
    \grmor w.l
    \grmor w[e]
    \\
    &\text{Statements}&
    S \grmeq &
    w = e
    \grmor \ifs eSS
    \grmor \whiles eeS
    \grmor \returns e
    \grmor S ; S
    \grmor \Empty
    \\
    &\text{Declarations}&
    D \grmeq & 
    \importdecl su \grmor
    \vardecl Txe \grmor
    \fundecl T x {\overline{T\,x}} {\overline{T\,x=e}; S}
    \\
    &\text{Programs}&
    P \grmeq & D ; P \grmor \Empty 
\end{align*}
  \caption{The syntax of \saferestscriptlang (extends Figure~\ref{fig:hr_language_syntax})}
  \label{fig:saferestscript}
\end{figure}


\subsection{Syntax} 

The syntax of \saferestscriptlang, presented in Figure~\ref{fig:saferestscript},
extends that of \headrestlang in Figure~\ref{fig:hr_language_syntax}.
%
%
The language includes a new \emph{constant} $\undefinedk$. 
Functions that return $\undefinedk$ are of type $\voidk$, an abbreviation of
$(\wheret{x}{\anyk}{x==\undefinedk})$.


At the level of \emph{expressions}, \saferestscriptlang introduces
support for 
 \rest calls $m\,u\,e$, composed of an HTTP method $m$ (see
Figure~\ref{fig:hr_language_syntax}), an URI template literal $u$
describing the relative URL of the target resource, and an expression
$e$ that should evaluate to a value of type $\Requestk$.
The endpoint needs to be specified in the
\saferestscriptlang specification imported by the program. 
%
Functions can be declared with $\asynck$ keyword; 
calls to these functions are asynchronous while  
\rest calls are asynchronous if they are preceded by keyword
$\awaitk$. 
Keywords $\asynck$ and $\awaitk$ are optional, as indicated
by the question mark in the syntax in Figure~\ref{fig:saferestscript}.
%


\emph{Statements} include variable assignment. The left hand side~$w$ of an assignment statement (a \emph{location})
is a variable~$x$, an object field $w.l$, or a position in an array
$w[e]$. An assignment can thus update a
specific element of an object or an array.
%
%
Moreover, statements include conditional statements,
while loops, and return statements.  Loops may declare an
invariant, i.e., an expression that is true at loop entry and after
each loop iteration. Invariants are sometimes necessary to prove that
certain expressions have the right type, for instance, whether the
effective type of the expression used in a return statement matches
the return type of the function.  Statement $\returnk$ abbreviates
$\returns\undefinedk$.


An \saferestscriptlang \emph{program} is a sequence of
\emph{declarations}: import clauses, global variable and function
declarations. The implementation of \saferestscriptlang further
supports type abbreviations in the form of $\typedecl xT$.
Global variables are introduced as $\vardecl Txe$.
Function definitions 
of the form $\fundecl T f {\overline{U\,x}} {\overline{V\,y=E}; S}$ 
are composed of a return type~$T$, the function name~$f$, a comma-separated list of
parameters with their respective types $\overline{U\,x}$, and the
function body. In order to simplify variable scope validation, the
body opens with the declaration and initialization of all local
variables: $\overline{V\,y=e}$ is a semi-colon-separated list of
variable declarations. The initialization is mandatory since some
types, such as refinement types, may not have a default value. 
The function's body consists of a statement $S$ that defines
the control flow and the return value.
For functions to be treated asynchronously they must be declared as
$\asynck$.
 Functions that call other asynchronous function must also be
 declared as $\asynck$.

\subsection{Type Checking}  

Statically type checking
\saferestscriptlang programs is a major challenge given the rich type
system of \saferestscriptlang and global imperative variables. It
requires flow-sensitive typing (the effective type of an expression
depends on its position in the control flow).

\saferestscriptlang programs are translated into verification conditions, i.e., logical formulas
whose validity entails the correctness of the program. Following a popular approach initiated by Spec\#~\cite{SpecSharp}, these conditions are not generated directly but instead obtained through a  translation into \boogielang \cite{Leino:2005}, an intermediate language for program verification. 
Once a \saferestscriptlang program is translated into a \boogielang program, it is up to the \boogielang validator to generate the verification conditions and, resorting to an SMT solver, verify whether they hold.


At the basis of the translation is an axiomatization of the typing
relation that is inspired by \whileylang~\cite{Pearce:2013}.
Values and types are modelled as sets. All \saferestscriptlang values,
independently of their type, belong to the \boogielang type 
\boogie{Value}. For each type, we introduce functions and axioms that
define its subset of values. 
More concretely, given a type \boogie{X} (for example, \safescript{Integer})  and its
internal representation \boogie{Y} in \boogielang (\boogie{int}, in the example), the base functions and axioms are
the following.

\begin{lstlisting}[language=boogie]
function isX(Value) returns (bool);
function toX(Value) returns (Y);
function fromX(Y) returns (Value);
axiom (forall y: Y :: isX(fromX(y)));
axiom (forall y: Y :: toX(fromX(y)) == y);
axiom (forall v: Value :: isX(v) ==> fromX(toX(v)) == v);
\end{lstlisting}

Function \boogie{isX} checks whether a value belongs to type
\boogie{X} and returns a \boogielang boolean. Function
\boogie{toX} converts the \boogielang value to its internal representation
\boogie{Y}, and \boogie{fromX} performs the inverse operation. The
axioms define the properties of the functions. 
The first asserts
that all values constructed from type \boogie{Y} belong to type
\boogie{X}. The second and third axioms assert that \boogie{toX} and
\boogie{fromX} are inverse functions.
More complex types, such as arrays and objects, are represented by \boogielang maps
and require the introduction of additional functions and axioms.    

Functions \boogie{isX}, \boogie{toX} and \boogie{fromX} are used for
defining the translation of expressions and the predicate  that checks
whether the value of an expression is of a given type. This is illustrated below in simple cases: the translation  of an  \saferestscriptlang integer literal and an array access, and the predicate for the integer type, the object type and the refinement type.

\begin{footnotesize}
    \begin{align*}
        \V{n}  &= \fromInt{n}  \\
        \V{e_1\,[e_2]} & = \getIndexValue{\V{e_1}}{\toInt{\V{e_2}}} \\
         \F{\keyword{int}}{e} &= \isInt e  \\
         \F{\{l\colon T\}}{e}  & = \isObject e \land \hasField el \land  \F{T}{\getFieldValue e l} \\
        \F{(\wheret xTe_1)}{e} & = \F Te \land \V{[e/x]e_1} == \V\truek
     \end{align*}
     \end{footnotesize}

The translation of \LangSRS to \boogielang is based on the collection of functions presented below.  We discuss some cases that convey the main ideas of how the translation works. The full set of rules is available in the appendix.

\begin{align*}
    \V e & \;\equiv\; \text{\boogielang expression of type \boogie{Value} that represents expression $e$}
    \\
    \Fnonopt T & \;\equiv\; \text{\boogielang predicate that checks
                 whether an expression is of type $T$}
    \\
    \VV ex & \;\equiv\; \text{Sequence of \boogielang statements that validates expression $e$} 
    \\
     & \; \; \; \;  \; \; \; \text{and places the corresponding Boogie expression in variable $x$}
    \\
    \W T & \;\equiv\; \text{Sequence of \boogielang statements that validates type $T$}
    \\
    \Us we & \;\equiv\; \text{Sequence of \boogielang statements that validates
             assignment $w=e$}
    \\
    \B S & \;\equiv\; \text{\boogielang statement that represents
           statement $S$}
    \\
    \B D & \;\equiv\; \text{\boogielang declaration that represents
           declaration $D$}
\end{align*}

The translation of \rest calls and of specification triples to \boogielang are the most interesting elements of the translation, as they accomplish the view of endpoints-as-functions discussed before. 
\rest calls are translated to  \boogielang using a function, named $\keyword{restCall}$, that receives as parameters the \rest method, the translation of a string $u'$ representing the URI template relative path $u$,
and the request object, and returns the response object.  
Each specification triple is translated into an axiom relating the return value of $\keyword{restCall}$ with the request call argument as follows.
\begin{align*}
  \B{\triple{e_1}mu{e_2}} =\;&
  \textbf{axiom} \; (\textbf{forall} \; \keyword{request} \colon \Value,\, \keyword{response} \colon \Value ::\\
    &\qquad\qquad\quad \restCall{m}{\V{u^\prime}}{\keyword{request}} == \keyword{response}\, \land\\
    &\qquad\qquad\quad \V{e_{1}} == \V{\truek} \Rightarrow \V{e_{2}} == \V{\truek})
\end{align*}
The translation of \rest calls is defined by the following rules:
\begin{align*}
    \VV{m\,u\,e}{x} =\;&  \VV{e}{y} ; \; \textbf{assert} \, \F{\keyword{Request}}{y};
      \\
     & x := \restCall{m}{\V u}{y};  \;  \textbf{assume} \, \F{\keyword{Response}}{x}
    \\
    \VV{\awaitk\,m\,u\,e}{x} =\;& \VV{m\,u\,e}{x};  \;  \textbf{havoc} \, g_1,\dots,g_p
\end{align*}
Expression $e$ in a synchronous REST call is validated and placed in a fresh
variable $y$.  Then an \textbf{assert} checks whether $y$ is of type
$\Requestk$. Function $\keyword{restCall}$ is called and its response
is stored in variable $x$. The response is assumed to be of type
$\Responsek$. Note that when the request does not meet the
pre-condition of any triple for the target endpoint, the
axiomatization of $\keyword{restCall}$ does not ensure anything about
the response; it is only known that it belongs to the $\Responsek$ type.
In asynchronous REST calls, the execution is suspended and, when
resumed, the global variables may have changed. This is captured by
the \textbf{havoc} statement, which assigns them arbitrary values
(respecting the declaration types).

The  type validation takes into account that types may contain
expressions by descending the abstract syntax tree of types. 
The most important rule is the rule for $\wherek$ types ($y,z$ are variables taken freshly).
\begin{equation*}
  \W{(\wheret xTe)}  = \W T; \textbf{assume} \, \F{T}{y}; \VV{e\subs{y}{x}}{z};
  \textbf{assert} \, \F{\booleank}{z}
\end{equation*}
We complete this brief presentation by addressing the translation of
global variable declaration and functions. 
In the first rule, the declared type is captured by a Boogie
\textbf{where} clause while the initialization is ignored as it is not
relevant:
whenever a procedure is called, nothing can be assumed
about any global variable besides its declaration type. The validation of $T$ and $e$ is achieved via an additional, dummy, procedure~$f$.
\begin{align*} 
 & \B{T \; x = e} = 
        \textbf{var} \, x : \valuek \;  \textbf{where} \;  \F Tx;\V{T \; f () \; \{\, \returnk \, e \,\}}  \\
 & \B{T \; f \; ( \, T_1\,x_1, \dots, T_n\,x_n \,) \; \{\, U_1 \; y_1=e_1 \,; \,\dots\, ; \, U_m \; y_m=e_m \,; \, S \,\}} = \\
  & \textbf{procedure} \; f (x_1: \Value, \dots, x_n: \Value) \; \textbf{returns} \; (\keyword{result}: \Value) \\
  & \qquad \textbf{requires} \; \F{T_1}{x_1} \land \dots \land \F{T_n}{x_n};  
      \; \;\textbf{ensures} \; \F{T}{\keyword{result}}; \\
  & \qquad \textbf{modifies} \; g_1, \dots, g_p;  \\
  & \{\\
  & \qquad \textbf{var} \, y_1, \dots, y_m, \, w_1, \dots, w_n : \Value; 
  w_1 := x_1; \dots; w_n := x_n; \\
  & \qquad \W{T_1}; \dots; \W{T_n} ; \W{U_1}; \dots; \W{U_m} ; \W{T} \\
  & \qquad \B {y_1=e_1 \,; \,\dots\, ; \, y_m=e_m \,;
    S\subs{w_1}{x_1}\dots\subs{w_n}{x_n} ; \returnk}\\
  &\}
\end{align*}
%
In the second rule, the immutability of procedures parameters in  \boogielang 
requires the declaration of new variables to use instead of the parameters 
in the function body. 
The \textbf{requires} clause checks whether the arguments
belong to the parameters types and the
\textbf{ensures} clause checks whether in all returning points of the
procedure the result of the function matches the function type. 
The \textbf{modifies} clause asserts
that all global variables can be modified by the procedure. The body of the procedure 
makes the validation of parameters types $T_i$, local variables types $U_k$ and 
the return type $T$.
The validation order allows that the validity of $T$ and  $U_k$ 
depend on $T_i$ and the validity of $T_i$
depend on the $T_j$, for $j<i$, as in 
 $\{\wheret{x}{\intk}{x>b/a}\} \; f(\{\wheret{x}{\intk}{x!=0}\}  \; a, \intk \; b)\{... \}$. 
Similarly to
\saferestscriptlang, \boogielang local variables must be declared at
the beginning of a procedure, so all fresh variables required by the
translation must also be declared before any \boogielang statement.

\subsection{Transpiling to  \javascriptlang} 

Valid programs are transpiled to \javascriptlang. 
The translation of \rest calls is achieved by calling auxiliary functions, one for synchronous and another by asynchronous calls. The URL to the call is the expansion of the URI template; its parameters are defined by the field template of the request object. The expansion follows the RFC 6570~\cite{rfc6570}, only for the level of URI templates supported by \LangSRS. The content-type JSON is added to the request headers, so objects sent and received in the body are ensured to be of JSON format, and therefore having a direct translation to \javascriptlang objects. The calls use \javascript{XMLHttpRequest}, an object that is supported by all browsers and devices.

Many REST APIs endpoints can only be used successfully with authentication, that is including a special token in the header authorization. The simpler and most common type of authentication is the Basic authentication scheme, defined by RFC 7617.
This authentication is not secure and must be used with HTTPS. The token must be encoded in base64, according to RFC 4648,
which is done by the native JavaScript function \textit{btoa}. To simply this process, an additional header \textit{basicAuthorization} is supported, that, before the call, performs the necessary encoding and adds the necessary authorization header.

The \saferestscript{await} and \saferestscript{async} keywords are directly translated to JavaScript. Functions that are asynchronous return a promise with the returning value, and the \javascript{await} operator reads that value.


\section{Evaluation}
\label{sec:evaluation}

This section addresses the evaluation of our approach. Ideally, we
would like to compare the bug finding efficacy of our approach in
``real code'' with that of Wittern at al.~\cite{Wittern:2017}, the unique approach to statically checking \rest calls  that we are aware of. However, this was not really feasible, as to translate JavaScript code into \LangSRS, requires to annotate the libraries used in the code and/or write adaptors that monitor the interface with those libraries.

In this way, to evaluate our approach, we used \headrestlang to
specify a variety of REST APIs and \LangSRS to write and validate
programs that exercise the different elements of the language while
consuming REST APIs. The goal is to evaluate {\it to what degree can
  \LangSRS be used in examples which include complex REST calls that
  can be found in real examples}.

Additionally, since  \LangSRS validator can be used for code that does not make any \rest call, we also evaluate {\it to what degree is the \LangSRS validator able to meet well known verification challenges and how does its performance compare with similar verification tools}.

In the next paragraphs we provide details about the specifications and programs we have developed, publicly available for download\footnote{\url{http://rss.di.fc.ul.pt/tools/confident/}}. 

\subsection{Verification Challenges}
\label{sec:vc}

In 2008, Wide et al~\cite{Weide:2008} proposed a suite composed of eight
incremental benchmarks for verification tools and techniques that
prove total correction of object-based and object-oriented software.
We selected the first two verification problems in this suit, since
the remaining ones were designed with inductive datatypes in mind
(e.g., breadth search in trees or find in linked lists) and, hence,
can not be addressed in \LangSRS. We then searched for more problems
in previous verification competitions that could be addressed by
\LangSRS and collected six additional problems: problems 1 and 2 from
VSComp2010~\cite{VSComp2010}, challenge 1 and 3 of COST Verification
Competition 2011~\cite{foveoos11},
and problems 1 and 3 from
VSTTE2012\footnote{\url{https://sites.google.com/site/vstte2012/compet}}. In
all four challenges we selected \emph{all} problems that could be
conceptually addressed by \saferestscriptlang.
\begin{enumerate}
\item[\#1] Adding and Multiplying Numbers. \emph{Verify an operation that adds two numbers by repeated incrementing. Verify an operation that multiplies two numbers by repeated addition, using the first operation to do the addition. Make one algorithm iterative, the other recursive.}

\item[\#2] Binary Search in an Array. \emph{Verify an operation that uses binary search to find a given entry in an array of entries that are in sorted order.}


\item[ \#3] Sum and Max. \emph{Compute the sum and max of the elements of an array and prove that the sum is less or equal that max times the length of the array.} 

\item[\#4] Inverse. \emph{Compute the inverse B of a sequence A and prove that B is indeed an inverse of A.} 

\item[\#5] Two-way Sort. \emph{Two-way sort algorithm of a boolean array, which sorts the array using only swaps. Prove that the array is sorted in increasing order and is a permutation of its initial contents.}

\item[ \#6] Max Array. \emph{Given a non-empty integer array, verify that the index returned by a given algorithm points to an element maximal in the array.}

\item[ \#7] Two Duplicates. \emph{Given an integer array of length greater or equal than 2 with at
least two duplicates, find such two values}. 

\item[ \#8] Ring Buffer. \emph{Implement a queue data structure using a ring buffer and prove safety and a test harness.} 
\end{enumerate}

For comparing the performance of the \LangSRS validator against
similar verification tools, we choose the \dafnylang
verifier~\cite{Leino10} since it also uses \boogielang as an
intermediate language, also relies on the Z3 SMT solver for
discharging proof obligations, and offers the option of compiling to
\javascriptlang. Moreover, \dafnylang solutions for all problems
under consideration are available at the language's GitHub
repository\footnote{\url{https://github.com/dafny-lang/dafny}}.

We were able to solve 6 out of the 8 problems (75\%): \#1--\#3, \#5,
\#6 and \#8. The solutions in \LangSRS for all but problem \#8 are
similar to those available for \dafnylang, with refinement types for
the I/O behaviour of functions capturing what in \dafnylang is
expressed through method contracts.  Our solution to problem \#5
returns the sorted array rather than performing sort in place, since
\LangSRS does not support references. In problem \#6 we used an
additional variable solely for the purpose of reasoning about the
program. In \dafnylang, these variables can be declared as ghost
variables (so that they do not play any role at run time), but this is
not the case in \LangSRS.  Our solution to problem \#8 is quite different
from that available for \dafnylang, which is based on the definition
of a class Queue implemented with a ring buffer. In \LangSRS, the
underlying data structure is defined as a type and queue operations
are implemented as functions over elements of this type.
 \begin{lstlisting}[language=saferestscript]
Type RingBuffer = b: {
  data: Integer[],    // buffer contents
  sizeof: Positive,   // buffer capacity
  ?first: Integer,    // head of queue, if any
  len: Natural	      // queue length
} where
  (b.len > 0 ==> (b in {first: Natural} && b.first < length(b.data))) && 
  b.sizeof == length(b.data)
\end{lstlisting}

We were able to validate the test harness, with assertions concerning
the head of a queue after several enqueue and dequeue operations. The
FIFO semantics of the queue, which is expressed as a Queue invariant
in the \dafnylang solution, is not expressible in \LangSRS.
On what concerns problems \#4 and \#7, we were not able to find satisfactory solutions. In both cases, there are some important properties that depend on loop invariants that the \LangSRS validator is not able to prove (even increasing the timeout, the validator does not reach any conclusion).  

\begin{table}[t]
\centering
\begin{tabular}{l|c|c|c|c|}
& \multicolumn{2}{c|}{\dafnylang v2.3.0} & \multicolumn{2}{c|}{\LangSRS} \\ \cline{2-5} 
& end-to-end & \boogielang only & end-to-end & \boogielang only\\ \hline
Adding and Multiplying & 1.9 & 1.0 & 0.7 & 0.6 \\
Binary Search & 1.9 & 1.0 & 2.8 & 2.7 \\
Sum Max & 1.9 & 1.1 & 0.6 & 0.5 \\
Two-Way Sort & 2.0 & 1.3 & 1.4 & 1.2 \\
Max Array & 1.8 & 1.0 & 1.0 & 0.9 \\
Ring Buffer & 3.1 & 2.4 & 1.6 & 1.4 \\ \hline
\end{tabular}
\caption{Comparison between \dafnylang and \LangSRS execution time (in seconds)} 
\label{tab:dafny_safescript}
\end{table}


Table~\ref{tab:dafny_safescript} presents a comparison of the
execution times of \dafnylang and \LangSRS. Benchmarking was performed
on a machine with an Intel Core i7-7700HQ CPU, with 2.80 GHz and 16 GB
of RAM memory, under Windows 10.  The two verifiers were run under the
same versions of \boogielang and Z3; the times
reported are the average of three runs.
The table shows (i) the execution time of each tool running from a
command line (\emph{end-to-end}), which includes the parsing,
validation and translation to \javascriptlang, and (ii) the execution
time of the validation step, which is achieved, in both cases, by
executing the \boogielang validator over the generated \boogielang
code. The performance of the two tools is relatively similar, with
\LangSRS being faster in all but one case. The validation of the  
\boogielang programs generated by \dafnylang takes 36\% longer on average.

\subsection{Consuming REST APIs } 

We used \LangSRS to write programs that consume publicly available
APIs and do not require authentication, including
PetStore\footnote{\url{https://petstore.swagger.io}} and
DummyAPI\footnote{\url{http://dummy.restapiexample.com}} as well as
programs that consume real-world off-the-shelf services such as
Instagram, GitHub and GitLab. Since API calls in \LangSRS are checked
against \headrestlang specifications, we had also to develop
\headrestlang specifications for the chosen APIs describing the
behaviour of the relevant endpoints. In what follows we provide
 details about three of these case studies (additional information in \cite{srsExamples}).

\paragraph{Instagram}

We developed in \LangSRS a solution alternative to the \javascriptlang
function in Section~\ref{sec:saferestscript-motivation}. The
application allows users to find Instagram photos by tag or location
and calls the different endpoints of the Instagram API which supports
search for (i)~locations by geographic coordinate, (ii)~photos by
location and (iii)~photos by tag.  The solution is based on a \LangSRS
program defining asynchronous functions for calling the API, similar
to that presented in Figure~\ref{fig:InstaClient}. These functions are
available in the generated \javascriptlang code and used by the
program that manipulates the DOM.  We additionally developed a
program for showing the recent comments on media for a user, given
its identifier, which requires to call three other endpoints: one to
get the ids of recent media, another to get the comments for each of
them and a third one to get information about the user.  Both programs
use the same  specification with the behaviour of
the 6 endpoints.



\paragraph{GitHub} 
We developed an \LangSRS program that offers a function
\saferestscript{getUserById(int id)} to obtain a GitHub user given its
\saferestscript{id} with return type
\saferestscript{(u: User where u.id == id)|NotFoundError}.  Since the
GitHub REST API does not have an endpoint that supports this operation
(to get the representation of individual users, one needs to provide
the username), our program sends a \saferestscript{GET} request to
\headrest|/users?since=id-1| if \headrest|id| is a positive integer.
According to the API documentation, this endpoint lists all users, in
the order that they signed up on GitHub. This is done by pagination:
each call retrieves a sublist of all users. The start of the sublist
is defined by the optional parameter \headrest{since}. If case the
parameter is not present, then its value is assumed to be zero.  In
the \headrestlang specification of GitHub we were able to precisely
express this behaviour. One of the assertions included in the
specification states that if the request provides a value for
\headrest|since| that is a natural number, then the array of users
provided in the response body starts with a user whose \headrest{id}
is equal to \headrest{since+1}.
%
%
This assertion is essential to prove that if a user is obtained, it has the id provided in the function argument and, hence, that the return type of the function is valid.

The endpoint \headrest|/get users{?since}| can also be used for
searching for an user with certain characteristics. We used it to
define a function \saferestscript{getSiteAdmin()} with return type
\saferestscript{(u: User where u.site_admin)|AdminNotFound} that
searches over the GitHub users to find an
administrator. 
The search code gets the various pages of users and stops when one of
them contains an user with admin privileges, or when no admin was
found on all pages, in which case an error message is returned. The
fact that the function type checks ensures that the returned user
representation (if any) is an indeed an administrator.

In GitHub each user has a set of repositories, and each repository has
a set of collaborators and a list of commits, each with its author. We
programmed a function that gets the collaborators of a repository that
did not contribute to a project, i.e., did not make a commit.  The
function crosses the information obtained in two different endpoints:
one
 \headrest|/repos/{owner}/{repo}/collaborators|, 
for retrieving the collaborators of a given repository 
\headrest{repo} 
for a given user 
\headrest{owner} 
and  another
\headrest|/repos/{owner}/{repo}/commits|, 
for retrieving the list of commits of the repository. As the repository may be private, the function receives a key that must give authorization to access the repository information, and that is added to the request header. 
 
\paragraph{GitLab}
is the Git manager that our students use to develop their
course projects. We used \LangSRS to program functions that automate
tasks we recurrently perform manually. For instance, we programmed a
function to remove a user from all projects owned by another
user. This function uses three endpoints, one of them involving
request and response types particularly large---the request has more
than 10 optional parameters and the response body is an array of an
object type with more than 30 fields, several of them also
objects. The type of the response is used, for instance, to validate
the expression \saferestscript|response.body[i].namespace.name| that
occurs in the body of the function.  Another interesting example is
the function
\saferestscript{getWikisFromProject(string token, int|string id, boolean withContent)}
we defined in \LangSRS to get the wikis from a project, identified by
its integer id or a string that is the URL-encoded path of the project
(and, hence, of type \saferestscript{int|string}). This function uses
an endpoint that has an optional parameter to indicate whether the
response should have the content of the wikis. The behaviour of GitLab
at this endpoint was specified in \headrestlang as shown below and
allows \LangSRS validator to find errors in accesses to
\saferestscript|response.body[i].content|, e.g., if the code does not
guarantee that such access is performed only if the value sent in
\saferestscript|request.template.with_content| is true.

 \begin{footnotesize}
 \begin{lstlisting}[language=headrest]
{ request in {template: {id: String|Integer, ?with_content: Boolean}} }
       get `/projects/{id}/wikis{?private_token,with_content}`
{ (response.code == 200 ==> response in {body: Wiki[]}) &&
  (response.code == 200 && request.template in {with_content: Boolean}  
    && request.template.with_content ==> 
    	response.body in (Wiki & {content:String})[]) }
\end{lstlisting}
\end{footnotesize}

\begin{table}[t]
\centering
\begin{tabular}{l|c|c|c|c|c|c|c|c|}
& \multicolumn{4}{c|}{\headrestlang} & \multicolumn{4}{c|}{\LangSRS} \\ \cline{2-9} 
&\#EndP    & \#Types & LOC   &  Check (s) & \#EndP      & \#Func & LOC  & Check (s)     \\ \hline  
Instagram\#app1       & 6          & 9  & 225       &  1.3 & 3             & 4 & 82    &   1.5    \\
Instagram\#app2       & 6      & 9      & 225        & 1.3 & 3             & 3 & 65     &  1.8   \\
GitHub & 5       & 9    & 93        & 3        & 0.8    &  3 & 86      & 1.3    \\
GitLab  & 10    & 20        & 435        & 8    &1.7       &  10 & 250    & 50.5    \\   \hline
\end{tabular}
\caption{Case studies of consuming REST APIs with \LangSRS } 
\label{tab:srs_examples}
\vspace{-2 em}
\end{table}

Table~\ref{tab:srs_examples} presents additional information
about the 3 case studies.
The first group of columns addresses \headrestlang specifications and
shows the number of endpoints that were specified, the number of types
that were defined, the number of lines and the validation time, in
seconds. The second group of columns, which addresses \LangSRS client
programs, shows the number of endpoints that were consumed, the number
of functions that were defined, the number of lines of code and the
validation time, in seconds. The validation the time of \LangSRS
programs presented in the table does not consider validation time of
the specification, since this step does not need to be repeated (unless the specification changes). 
The time benchmarks were taken in the same conditions described in Section~\ref{sec:vc}. 

\medskip

Overall, these examples demonstrate that \headrestlang supports the
specification of a variety of API endpoints found in real examples and
is able to capture important properties of these endpoints that were
previously available only in natural language. During the development
of the client programs we could witness that the formalisation of
properties allowed \LangSRS to find all sort of errors in our code, in
particular, errors in the invocations of the underlying services
(invalid or missing data in the requests or use of incorrect URLs) and
errors in the use of the data received in the response. We also noted that, 
if we were programming these client programs in \javascriptlang, most of the 
errors we made would not be found by the tool presented in \cite{Wittern:2017}. 
On the one hand, errors caused by invalid data in the requests were often caused by 
restrictions on data that are simply not expressible in OpenAPI. On the other hand, 
several errors were in the use of the data received in the response, a type of error 
that is not addressed by the analysis performed by the tool.

In terms of
performance, we witness what is also evident in the results in
Table~\ref{tab:srs_examples}: the complexity of the types involved in
REST calls significantly slows the validation process when the
correctness of the code strongly depends on these types. This problem
can be alleviated by placing functions whose validation is too
demanding in separated source files. Because the validator ignores 
files that have not changed, these functions do not need to be validated 
again if they have not changed.


\section{Related Work}
\label{sec:relatedwork}

Static verification of \javascriptlang code has been the main research
topic for client-side coding in the last few years~\cite{Sun:2017}.
Neverthless, research concerning the verification of consumer code of
REST APIs for \javascriptlang-like client-side languages is
slim and the solutions proposed tend to be quite limited.
%
%

Solutions for helping finding bugs in scripts come in the form of a varied set of languages and tools. 
JSHint~\cite{JSHint} scans \javascriptlang code for suspicious usage;
Thiemann~\cite{Thiemann:2005} and Anderson et
al.~\cite{Anderson:2005} propose type system for subsets of
\javascriptlang;
\typescriptlang~\cite{Bierman:2014}, Dart~\cite{Dart}, and
Flow~\cite{Flow} are languages that were developed with the goal of
statically detecting type-related errors in \javascriptlang-like
languages.
Languages such as Dependent
JavaScript~\cite{DBLP:conf/oopsla/ChughHJ12} and Refined
TypeScript~\cite{Vekris:2016} incorporate sophisticated type systems,
but the power of the $\inexp$ predicate and semantic subtyping
(supported by \saferestscriptlang) seems to be particularly suited for
programming REST clients.
%
%
%
\whileylang~\cite{Pearce:2013} is a 
programming language that features a rich type system and flow typing;
it uses \boogielang only to check the verification conditions~\cite{Utting:2017}.
Contrary to \saferestscriptlang, neither of these solutions
specifically addresses REST calls.


$\typescriptlang_{\text{IPC}}$~\cite{DBLP:conf/ecoop/OostvogelsKM18}
extends \typescriptlang with the ability to describe the presence or
absence of properties in objects, a feature that \headrestlang and
\saferestscriptlang can easily describe and for which a derived
predicate $\isdefinedk$ was introduced
(cf.~Section~\ref{sec:headrest}). Like all the languages discussed
above, $\typescriptlang_{\text{IPC}}$ does not provide explicit
support for REST calls.

The tool by Wittern et al.~\cite{Wittern:2017}, discussed in the
introduction, statically checks web API requests in \javascriptlang
code, focusing on ajax requests made via jQuery\footnote{\url{https://api.jquery.com}}. The
tool uses a field-based call graph to make the necessary string
analyses on the \javascriptlang method calls and is able to check
whether calls to endpoints match a valid URI template in the API
specification and the request has the expected data.
Such errors are easier to check in \saferestscriptlang since the
construction of URIs is limited to URI template instantiation (thus
ruling out the construction of new URIs via string operations such as
concatenation). In contrast, the verification supported by
\saferestscriptlang that the request has the expected data is beyond
reach of Wittern et al.\ for the rich, non-OpenAPI, data definitions.
RESTyped Axios~\cite{Dezfuli-Arjomandi:2017} is a client-side tool
that verifies REST calls in \typescriptlang against RESTyped specifications, with requests made via the Axios framework~\cite{axios}. RESTyped allows to define strongly-typed routes and Axios checks at compile time whether
the URLs are valid and whether the types of the members passed on
requests and accessed on responses correspond to the ones declared in
the specification.
These two approaches fail to detect many of the defects at the reach
of \saferestscriptlang, including those related to complex
restrictions on input data of REST calls (not expressible in the
adopted specification languages) or the misuse of the return data.

\whiplang~\cite{WayeCD17} is a contract system for services that uses
a dependent type system to monitor services at run-time and check
whether they respect their advertised interfaces. \whiplang offers a
high-order contract language that, similarly to \headrestlang,
addresses the lack of expressiveness of IDLs to capture non-trivial
properties that can be found in the documentation of popular services.
\whiplang focus is on the
specification of properties that cross-cut more than one service (e.g., properties that describe how a client of one system should
use a reply to interact with another) and, by using contracts,
addresses the specification of the expectations and promises of a
service to other services.


\section{Conclusion}
\label{sec:conclusion}

We present a framework for statically checking code that consumes
APIs. Relevant aspects of APIs are described with \headrestlang, a
specification language featuring refinement types and semantic
subtyping. The consumer code itself is written in \saferestscriptlang,
a variant of \javascriptlang with explicit primitives for synchronous
and asynchronous REST calls. \headrestlang specifications are
validated by resorting to an SMT solver to discharge semantic subtyping
goals. API consumer code is checked via a translation to \boogielang.
We validate our approach by writing in \saferestscriptlang various
benchmarks from the literature.
We further report on three case studies of consumer code
for popular APIs (Instagram, GitHub, and GitLab).

\paragraph{Future work.} Much remains to be done. We sketch a few ideas for future work.

\begin{description}
\item[References] The lack of references is the most relevant difference between \LangSRS and
\javascriptlang. Introducing references in objects and arrays is not
trivial and adds additional complexity to the \boogielang
translation. \dafnylang\cite{Leino10} devised a clever solution using object
references in its translation to \boogielang, but the technique does
not carry straightforwardly to refinement types. A preliminary
experience showed that this extension substantially increases the
validation time.

\item[Functions as values] Functions are values in \javascriptlang. They are particularly useful
 in \javascriptlang when used with asynchronous code, since callbacks
 are often used. Adding functions as values will make \LangSRS one step
 closer to \javascriptlang. We envisage no major difficulty in this
 extension.
\item[External functions]
Since \LangSRS compiles to \javascriptlang, programmers may take
advantage of its  
standard libraries. To use \javascriptlang
functions in \LangSRS code, their signatures are required. We plan
to address this issue, possibly by following the \typescriptlang
approach, that is, by introducing declaration files where external \javascriptlang 
can be declared and annotated with the \LangSRS types so they can be
used in \LangSRS code.
\item[Inconsistent specifications]
\LangHR specifications may feature inconsistent triples. This aspect
does not influence the validation of \LangHR specifications, since
each triple is validated independently, but it can affect the
validation of \LangSRS programs. Specifications featuring inconsistent
triples induce inconsistent \boogielang axiomatizations, allowing
programs with typing errors to be validated. It is therefore
important to detect inconsistent \LangHR specifications.

\item[URI templates] The relative URL endpoint for making a REST call must be a string literal, that matches an URI template in the imported \LangHR specification. Therefore, currently it is not possible to construct the URL string with the values of the URI template variables, but these must be indirectly indicated in the object request. With the introduction of UriTemplates as values, as in \LangHR, the REST calls may be simpler and more similar with what if found in \javascriptlang code.
\end{description}

\paragraph{Acknowledgements}

%
This work was supported by FCT through the LASIGE Research Unit, ref.\
UIDB/00408/2020, and by project Confident ref.\
PTDC/EEI-CTP/4503/2014.

\bibliographystyle{splncs04}
\bibliography{bibliography}

\newpage
\appendix
\section{URI Templates}
\label{sec:full-set-rules}

\begin{description}
\item[URI template syntax:] Figure~\ref{fig:hr_uri_template_syntax}.
  \begin{figure}[t]
  \begin{align*}
    &\text{URI Templates}& u \grmeq & \texttt{`}t\texttt{`}
    \\
    &\text{Terms}&
      t \grmeq & \Empty
      \grmor l \, t
      \grmor \{v, \dots, v\} \, t
      \grmor \{?v, \dots, v\} \, t
    \\
    & \text{Literals}& l \grmeq & 
      \texttt{([\^{}"'\%<>\textbackslash\^{}`\{|\}] | \%[0-9A-F]\{2\})+}\\
    & \text{Variables}& v \grmeq &
     \texttt{([0-9A-Z\_a-z]|\%[0-9A-F]\{2\})}
     \texttt{([.0-9A-Z\_a-z]|\%[0-9A-F]\{2\})*}
  \end{align*}
  \caption{The syntax of URI templates}
  \label{fig:hr_uri_template_syntax}
\end{figure}


\item[URI template extraction:] Figure~\ref{fig:hr_uri_template_extraction}.
  \begin{figure}[t]
  \begin{gather*}
    \frac{
      \isUri tT
    }{
      \isUri{`t`} T
    }
    \quad
    \frac{
    }{
      \isUri{\Empty}{\{\}}
    }
    \quad
    \frac{
      \isUri tT
    }{
      \isUri{l \; t} T
    }
    \quad
    \frac{
      \isUri tT
    }{
      \isUri{\{?\overline v\} \, t}{T}
    }
    \quad
    \frac{
      \isUri tT
    }{
      \isUri{\{\overline v\} \, t}{T \& \{\templateid\colon\{\overline{v\colon\anyk}\}\}}
    }
  \end{gather*}
  \caption{URI template type extraction: $\isUri uT$}
  \label{fig:hr_uri_template_extraction}
\end{figure}


\end{description}


\section{The extraction functions}
\label{sec:extraction}

\begin{description}
\item[Normal disjunction:] Figure~\ref{fig:hr_dnf}.
\begin{figure}[t]
    \begin{align*}
      &\text{Normal disjunction}&
      D \grmeq& R_1 \vee\dots\vee R_n \qquad(n\ge0, \emptyk\text{ when }n=0)
      \\
      &\text{Normal refined conjunction}&
      R \grmeq& \wheret{x}{C}{e}
      \\
      &\text{Normal conjunction}&
      C \grmeq& A_1 \wedge \dots\wedge A_n \qquad(n\ge0, \anyk\text{ when }n=0)
      \\
      &\text{Atomic type}&
      A \grmeq& G  \grmor \alpha \grmor \{\} \grmor \objecttype[D] \grmor \arraytype[D]
    \end{align*}
    \caption{Disjunctive normal form types}
    \label{fig:hr_dnf}
  \end{figure}

  \item[Type normalisation:] Figure~\ref{fig:hr_normalisation}.
  \begin{figure}[t]
    \begin{align*}
      & \norm \anyk \eqdef \anyk
      \\
      & \norm G \eqdef \wheret{y}{G}{\truek}
      \\
      & \norm \alpha \eqdef \wheret{y}{\alpha}{\truek}
      \\
      & \norm{\{\}} \eqdef \wheret{y}{\{\}}{\truek}
      \\
      & \norm{\{l\colon T\}} \eqdef \wheret{y}{\{l\colon T\}}{\truek}
     \\
      & \norm{\arraytype} \eqdef \wheret{y}{\arraytype}{\truek}
      \\
      & \norm{\wheret xTe} \eqdef \bigvee_{i=1}^n
                                 \conjDD{\wheret{x_i}{C_i}{e_i}, \normr{\wheret{x}{C_i}{e}}}
      \\
      & \qquad \text{if } \bigvee_{i=1}^n (\wheret{x_i}{C_i}{e_i}) = \normr{T}
      \\
      \\
     & \normr{\wheret{x}{C}{\isin xT}} \eqdef %
        \norm{C \AndT T}\qquad\text{if } x\notin\fv(T)
      \\
      & \normr{\wheret{x}{C}{e_1\OrE e_2}} \eqdef 
        \normr{\wheret{x}{C}{e_1}} \vee \normr{\wheret{x}{C}{e_2}} 
      \\
      & \normr{\wheret{x}{C}{e_1\AndE e_2}} \eqdef 
        \conjDD{\normr{\wheret{x}{C}{e_1}}, \normr{\wheret{x}{C}{e_2}}}
      \\
      & \normr{\wheret xCe} \eqdef \wheret xCe \qquad\text{otherwise}
      \\
      \\
      & \conjDD{(R_1 \vee\dots\vee R_n), D} \eqdef
        \conjRD{R_1,D} \vee\dots\vee \conjRD{R_n,D}
      \\
      & \conjRD{R, (R_1 \vee\dots\vee R_n)} \eqdef
        \conjRR{R,R_1} \vee\dots\vee \conjRR{R,R_n}
      \\
      & \conjRR{\wheret{x_1}{C_1}{e_1}, \wheret{x_2}{C_2}{e_2}} \eqdef
        \wheret{y}{C_1\wedge C_2}{\subs{y}{x_1}e_1 \AndE \subs{y}{x_2}e_2}
    \end{align*}
    where $y$ is a fresh variable in all cases
    \caption{Type normalisation: $\norm T = D$}
    \label{fig:hr_normalisation}
  \end{figure}

  \item[Extraction of object field type:] Figure~\ref{fig:hr_field_extraction}.
  \begin{figure}[t]
    \centering
    \setlength{\jot}{10pt}
    \begin{gather*}
      \frac{
          R_i.l \leadsto U_i
          \quad 
          \forall i\in 1..n
      }{
          (R_1 \, | \, \dots \, | \, R_n).l \leadsto (U_1 \, | \, \dots \, | \, U_n)
      }
      \tag{Field Disj}
      \\
      \frac{
          C.l \leadsto U
      }{
          (\wheret xCe).l \leadsto U
      }
      \tag{Field Refine}
      \\
      \frac{
          (S = \{U_i \, | \, A_i.l \leadsto U_i\}) \ne \emptyset
      }{
          (A_1 \, \& \, \dots \, \& \, A_n).l \leadsto (\& \, S)
      }
      \qquad
      \frac{}{
        \{l:T\}.l \leadsto T
      }
      \tag{Field Conj, Field Atom}
    \end{gather*}
    \caption{Extraction of type of a field in an object type : $D.l \leadsto U$}
    \label{fig:hr_field_extraction}
  \end{figure}

  \item[Extraction of array item type:] Figure~\ref{fig:hr_item_extraction}.
  \begin{figure}[t]
    \centering
    \setlength{\jot}{10pt}
    \begin{gather*}
      \frac{
          R_i.\itemsk \leadsto U_i
          \quad 
          \forall i\in 1..n
      }{
          (R_1 \, | \, \dots \, | \, R_n).\itemsk \leadsto (U_1 \, | \, \dots \, | \, U_n)
      }
      \tag{Items Disj}
      \\
      \frac{
          C.\itemsk \leadsto U
      }{
          (\wheret xCe).\itemsk \leadsto U
      }
      \tag{Items Refine}
      \\
      \frac{
          (S = \{U_i \, | \, A_i.\itemsk \leadsto U_i\}) \ne \emptyset
      }{
          (A_1 \, \& \, \dots \, \& \, A_n).\itemsk \leadsto (\& \, S)
      }
      \qquad
      \frac{}{
        \arraytype[T].\itemsk \leadsto T
      }
      \tag{Items Conj, Items Atom}
    \end{gather*}
    \caption{Extraction of the type of the items in an array type: $D.\itemsk \leadsto U$}
    \label{fig:hr_item_extraction}
  \end{figure}

\end{description}


\section{The Complete Translation of \saferestscriptlang to \boogielang}
\label{sec:boogie-complete}

\begin{description}
\item[Translation of expressions:] Figure~\ref{fig:ss_boogie_value}.
\begin{figure}[t]
    \centering
    \begin{align*}
        \V{\Forall xTe} & = \fromBool{(\textbf{forall} \; x\colon \Value :: \F Tx \Rightarrow \V e == \V\truek)} \\
        \V{\Exists xTe} & = \fromBool{(\textbf{exists} \; x\colon \Value :: \F Tx \land \V e == \V\truek)} \\
        \V{\ite{e_1}{e_2}{e_3}} & = \If \; \V{e_1} == \V\truek \; \Then \; \V{e_2} \; \Else \; \V{e_3} \\
        \V\inexp & = \fromBool{\F{T}{\V e}} \\
        \V{e_1\,[e_2]} & = \getIndexValue{\V{e_1}}{\toInt{\V{e_2}}} \\
        \V{e.l} & = \getFieldValue{\V{e}}{l} \\
        \V{f(e_1,\dots,e_n)} & = f(\V{e_1},\dots,\V{e_n}) \\
        \V{\{l_1\colon e_1,\dots,l_n\colon e_n\}} & = \keyword{fromObject}(\objectConst[l_1:=\maybeOf{\V{e_1}}] \\
        & \qquad \dots[l_n:=\maybeOf{\V{e_n}}]) \\
        \V{[e_1,\dots,e_n]} & = \keyword{fromArray}(\arrayConst[0:=\maybeOf{\V{e_1}}] \\
        & \qquad \dots[n-1:=\maybeOf{\V{e_n}}], n) \\
        \V{\truek} & = \true \\
        \V{\falsek} & = \false \\
        \V{n} & = \fromInt{n} \\
        \V{"c_1\dots\,c_n"} & = \fromString{\emptyString[0:=c_1]\dots[n-1:=c_n]\,}{n} \\
        \V{\nullk} & = \nullB \\
        \V{\undefinedk} & = \undefinedB \\
        \V{x} & = x
    \end{align*}
    \caption{Translation of expressions: $\V e$}
    \label{fig:ss_boogie_value}
\end{figure}

\item[Translation of types as predicates:] Figure~\ref{fig:ss_boogie_in_type}.
\begin{figure}[t]
    \centering
    \begin{align*}
        \F{\anyk}{e} & = \truek \\
        \F{\integerk}{e} & = \isInt e \\
        \F{\booleank}{e} & = \isBool e \\
        \F{\stringk}{e} & = \isString e \\
        \F{\arraytype}{e} & = \isArray e \, \land \, (\textbf{forall} \; y\colon \keyword{int} :: \isValidIndex{e}{y} \\ 
        & \qquad \Rightarrow \F {T}{\getIndexValue{e}{y}}) \\
        \F{\{\}}{e} & = \isObject e \\
        \F{\{l\colon T\}}{e} & = \isObject e \land \hasField el \land  \F{T}{\getFieldValue e l} \\
        \F{(\wheret xTe_1)}{e} & = \F Te \land \V{[e/x]e_1} == \V\truek
    \end{align*}
    where $y$ is a fresh variable
    \caption{Translation of types as predicates: $\F Te$}
    \label{fig:ss_boogie_in_type}
\end{figure}

\item[Translation of operator names:] Figure~\ref{fig:ss_boogie_operators_semantic}.
\begin{figure}[h!]
    \centering
    \begin{align*}
      \V{<=>} & = \keyword{equi} &
      \V{=>} & = \keyword{imp} &
      \V{|} & = \keyword{or} &
      \V{\&} & = \keyword{and} \\
      \V{==} & = \keyword{eq} &
      \V{!=} & = \keyword{ne} &
      \V{<} & = \keyword{lt} &
      \V{<=} & = \keyword{le} \\
      \V{>} & = \keyword{gt} &
      \V{>=} & = \keyword{ge} &
      \V{+} & = \keyword{sum} &
      \V{-} & = \keyword{sub} \\
      \V{++} & = \keyword{concat} &
      \V{*} & = \keyword{mult} &
      \V{\%} & = \keyword{rem} &
      \V{/} & = \keyword{div} \\
      \V{!} & = \keyword{neg} &
      \V{-} & = \keyword{min} &
      \V{\mkarrayk} & = \keyword{mkArray} &
      \V{\lengthk} & = \keyword{length} \\
      \V{\sizek} & =  \keyword{size} &
      &&
      &&
      &&
    \end{align*}
    \caption{Translation of operator names: $\V f$, where f is a primitive operator}
    \label{fig:ss_boogie_operators_semantic}
  \end{figure}

  \item[Translation of expressions with type validation:] Figure~\ref{fig:ss_boogie_value_validation}.
\begin{figure}[h!]
    \centering
    \begin{align*}
        \VV{\Forall {x_1}Te}{x} & = \W T \; \textbf{assume} \, \F{T}{y_1}; \; \VV{[y_1/x_1]e}{y_2} \\
        & \qquad \textbf{assert} \, \F{\keyword{boolean}}{y_2}; \\ 
        & \qquad x := \V{\Forall {x_1}{T}{x_1 == y_1 \Rightarrow y_2}}; \\
        \VV{\Exists {x_1}Te}{x} & = \W T \; \textbf{assume} \, \F{T}{y_1}; \; \VV{[y_1/x_1]e}{y_2} \\
        & \qquad \textbf{assert} \, \F{\keyword{boolean}}{y_2}; \\ 
        & \qquad x := \V{\Exists {x_1}{T}{x_1 == y_1 \land y_2}}; \\
        \VV{\ite{e_1}{e_2}{e_3}}{x} & = \VV{e_1}{y_1} \; \textbf{assert}\, \F{\keyword{boolean}}{y_1}; \\
        & \qquad \If\,(y_1 == \V\truek) \; \{\, \VV{e_2}{x} \,\} \; \Else \; \{\, \VV{e_3}{x} \,\} \\
        \VV{\oplus(e_1,\dots,e_n)}{x} & = \VV{e_1}{y_1} \dots \VV{e_n}{y_n} \\ 
        & \qquad \textbf{assert}\, \F{T_1}{y_1} \land \dots \land \F{T_n}{y_n}; \\
        \VV\inexp{x} & = \VV{e}{y} \; \W{T} \; x :=  \V {\isin{y}{T}}; \\
        \VV{e_1[e_2]}{x} & = \VV{e_1}{y_1} \; \VV{e_2}{y_2} \; \textbf{assert}\, \F{\arraytype[\anyk]}{y_1} \, \land \\ 
        & \qquad \F{(i: \keyword{Natural} \; \keyword{where} \; i < \lengthk(y_1)}{y_2}; \; x := \V{y_1[y_2]}; \\
        \VV{e.l}{x} & = \VV{e}{y} \; \textbf{assert} \, \F{\{l\colon \keyword{any}\}}{y}; \; x := \V{y.l}; \\
        \VV{f(e_1,\dots,e_n)}{x} & = \VV{e_1}{y_1} \dots \VV{e_n}{y_n} \\
        \text{if $f$ is a operator} & \qquad \textbf{assert} \, \F{T_1}{y_1} \land \dots \land \F{T_n}{y_n}; \\
        & \qquad x := \V{f(y_1,\dots,y_n)}; \; \text{where} \; f \colon (x_1\colon T_1),\dots,(x_n\colon T_n) \rightarrow T \\
        \VV{f(e_1,\dots,e_n)}{x} & = \VV{e_1}{y_1} \dots \VV{e_n}{y_n} \\
        \text{if $f$ not a operator} & \qquad \textbf{call} \; x := \V{f(y_1,\dots,y_n)}; \; \text{where} \; f \colon(x_1\colon T_1),\dots,(x_n\colon T_n) \rightarrow T \\
        \VV{\{l_1\colon e_1,\dots,l_n\colon e_n\}}{x} & = \VV{e_1}{y_1} \dots \VV{e_n}{y_n} \; x := \V{\{l_1\colon y_1,\dots,l_n\colon y_n\}}; \\
        \VV{[e_1,\dots,e_n]}{x} & = \VV{e_1}{y_1} \dots \VV{e_n}{y_n} \; x := \V{[y_1,\dots,y_n]}; \\
        \VV{c}{x} & = x := \V{c}; \\
        \VV{x}{z} & = z := \V{x};
    \end{align*}  
    where $y, y_1,\dots, y_n$ are fresh variables
    \caption{Translation of expressions with type validation: $\VV ex$}
    \label{fig:ss_boogie_value_validation}
\end{figure}

\item[Translation of type formation:] Figure~\ref{fig:ss_boogie_type_formation}.
\begin{figure}[t]
    \centering
    \begin{align*}
        \W{\arraytype} & = \W T \\
        \W{\{l\colon T\}} & = \W T \\
        \W{(\wheret xTe)} & = \W T ; \textbf{assume} \, \F{T}{y_1};  \VV{e\subs{y_1}{x}}{y_2};
         \textbf{assert} \, \F{\keyword{boolean}}{y_2}; \\
        \text{otherwise} \quad \W{T} & = \Empty
    \end{align*}
    where $y_1$ and $y_2$ are fresh variables
    \caption{Translation of type formation: $\W T$}
    \label{fig:ss_boogie_type_formation}
\end{figure}

\item[Translation of variable update:] Figure~\ref{fig:ss_boogie_variable_update}.
\begin{figure}[t]
    \centering
    \begin{align*}
        \Us{x}{e} & = x := e; \; \textbf{assert} \, \F{T_1}{z_1} \land \dots \land \F{T_n}{z_n}; \\
        \Us{u\,[e_1]}{e} & = \VV{u}{y_1} \; \VV{e_1}{y_2} \; \VV{y_1[y_2]}{y_3} \; \Us{u}{\arrayUpdate{y_1}{y_2}{e}}  \\
        \Us{u.l}{e} & = \VV{u}{y_1} \; \VV{y_1.l}{y_2} \; \Us{u}{\objectUpdate{y_1}{l}{e}}
    \end{align*}
    where $y_1, y_2$ and $y_3$ are fresh variables, $z_1, \dots, z_n$ the variables in scope, and $T_1, \dots, T_n$ types of those variables
    \caption{Translation of variable update: $\Us ue$}
    \label{fig:ss_boogie_variable_update}
\end{figure}

\item[Translation of statements:] Figure~\ref{fig:ss_boogie_statements}.
\begin{figure}[t]
    \centering
    \begin{align*}
        \B{u = e} & = \VV{e}{y} \; \Us{u}{y} \\
        \B{\ifk \; ( \, e \,) \; S_1 \; \elsek \; S_2} & = \VV{e}{y} \; \textbf{assert} \, \F{\booleank}{y}; \\
        & \qquad \If \; (y == \V{\truek}) \; \{\,\B{S_1}\,\} \; \Else \; \{\,\B{S_2}\,\} \\
        \B{\whilek \; ( \, e_1 \, ) \; \invk \, e_2 \; S} & = \VV{e_1}{y_1} \; \VV{e_2}{y_2} \\
        & \qquad \textbf{assert} \, \F{\booleank}{y_1} \land \F{\booleank}{y_2}; \\
        & \qquad \textbf{while} \, (\V{y_1} == \V{\truek}) \\
        & \qquad \textbf{invariant} \; \V{e_2} == \V{\truek}; \\
        & \qquad \textbf{free invariant} \; \F{T_1}{z_1} \land \dots \land \F{T_n}{z_n}; \\
        & \qquad \{\, \B{S} \; \VV{e_1}{y_1} \; \textbf{assert} \, \F{\booleank}{y_1};  \,\} \\
        \B{\returnk \; e} & = \VV{e}{\keyword{result}} \; \textbf{return}; \\
        \B{S_1;S_2} & = \B{S_1} \; \B{S_2} \\
        \B{\Empty} & = \Empty
    \end{align*}  
    where $y$, $y_1$, $y_2$ and $y_3$ are fresh variables, $z_1, \dots, z_n$ the variables in scope, and $T_1, \dots, T_n$ declared types of those variables
    \caption{Translation of statements: $\B S$}
    \label{fig:ss_boogie_statements}
\end{figure}

\item[Translation of function definitions:] Figure~\ref{fig:ss_boogie_function}.
\begin{figure}[t]
    \centering
    \begin{align*}
        &\B{T \; f \; ( \, T_1\,x_1, \dots, T_n\,x_n \,) \; \{\, U_1 \; y_1=e_1 \,; \,\dots\, ; \, U_m \; y_m=e_m \,; \, S \,\}} = \\
        & \textbf{procedure} \; f (x_1: \Value, \dots, x_n: \Value) \; \textbf{returns} \; (\keyword{result}: \Value) \\
        & \textbf{requires} \; \F{T_1}{x_1} \land \dots \land \F{T_n}{x_n}; \\
        & \textbf{modifies} \; g_1, \dots, g_p; \\
        & \textbf{ensures} \; \F{T}{\keyword{result}}; \; \\
        & \{ \, \textbf{var} \, z_1, \dots, z_k,\, y_1, \dots, y_m, \, w_1, \dots, w_n, : \Value; \\
        & w_1 := x_1; \dots\, w_n := x_n; \\
        & \W{T_1} \dots \W{T_n} \; \W{U_1} \dots \W{U_m} \; \W{T} \\
        & \B {y_1=e_1 \,; \,\dots\, ; \, y_m=e_m \,; [w_1/x_1]\dots[w_n/x_n]S \, ; \, \returnk} \, \}
    \end{align*}
    where $y_1,\dots,y_n$ are fresh variables representing each parameter, $z_1,\dots,z_k$ are the respective fresh local variables of the procedure, and $g_1,\dots,g_p$ are the global variables
    \caption{Translation of function definitions: $\B F$}
    \label{fig:ss_boogie_function}
\end{figure}
\end{description}







\end{document}
